\documentclass[aps,pre,showpacs,noshowkeys,amsmath,amssymb,amsfonts,superscriptaddress,longbibliography,reprint]{revtex4-1}
\usepackage[english]{babel}

\usepackage{graphicx}
\usepackage{bm}
\usepackage{physics}
\usepackage{mathtools}
\usepackage{gensymb}
\usepackage{multirow}

\setcitestyle{super}
\usepackage{caption}
\usepackage{subcaption}
\DeclareCaptionLabelSeparator{bar}{~\rule[-0.4ex]{0.2ex}{1em}~}
\DeclareCaptionLabelFormat{subfor}{\textbf{#2}}
\usepackage{xcolor}
\definecolor{UBcolor}{HTML}{007CC1}
\usepackage[colorlinks=true,pdfnewwindow=true,linkcolor=UBcolor,citecolor=UBcolor,urlcolor=UBcolor,breaklinks=true,linktocpage]{hyperref}
\usepackage[all]{hypcap}
\usepackage[nameinlink,capitalise]{cleveref}
\crefname{SI section}{SI Section}{SI Sections}
\Crefname{SI section}{SI Section}{SI Sections}
\begin{document}

\title{Collective durotaxis of cohesive cell clusters on a stiffness gradient}

\author{Irina Pi-Jaum\`a}
\affiliation{Departament de F\'{i­}sica de la Mat\`{e}ria Condensada, Universitat de Barcelona, Av. Diagonal 647, 08028 Barcelona, Spain}
\affiliation{Universitat de Barcelona Institut of Complex Systems (UBICS), 08028 Barcelona, Spain}

\author{Ricard Alert}
\affiliation{Princeton Center for Theoretical Science, Princeton University, Princeton, NJ 08544, USA}
\affiliation{Lewis-Sigler Institute for Integrative Genomics, Princeton University, Princeton, NJ 08544, USA}
\affiliation{Max Planck Institute for the Physics of Complex Systems, N\"{o}thnitzerst. 38, 01187 Dresden, Germany}
\affiliation{Center for Systems Biology Dresden, Pfotenhauerst. 108, 01307 Dresden, Germany}

\author{Jaume Casademunt}
\affiliation{Departament de F\'{i­}sica de la Mat\`{e}ria Condensada, Universitat de Barcelona, Av. Diagonal 647, 08028 Barcelona, Spain}
\affiliation{Universitat de Barcelona Institut of Complex Systems (UBICS), 08028 Barcelona, Spain}

\date{\today}

\begin{abstract}
Many types of motile cells perform durotaxis, namely, directed migration following gradients of substrate stiffness. Recent experiments have revealed that cell monolayers can migrate toward stiffer regions even when individual cells do not --- a phenomenon known as collective durotaxis. Here we address the spontaneous motion of finite cohesive cell monolayers on a stiffness gradient. We theoretically analyze a continuum active polar fluid model that has been tested in recent wetting assays of epithelial tissues,  and includes two types of active forces (cell-substrate traction and cell-cell contractility). The competition between the two active forces determines whether a cell monolayer spreads or contracts. Here, we show that this model generically predicts collective durotaxis, and that it features a variety of dynamical regimes as a result of the interplay between the spreading state and the global propagation, including sequential contraction and spreading of the monolayer as it moves toward higher stiffness. We solve the model exactly in some relevant cases, which provides both physical insights into the mechanisms of tissue durotaxis and spreading as well as a variety of predictions that could guide the design of future experiments.
\end{abstract}

\maketitle

\section{Introduction}
\label{intro}
The organized motion of cohesive groups of cells, usually referred to as collective cell migration,  plays a key role in many instances of morphogenesis, tissue regeneration and cancer invasion \cite{Friedl2009,Vedula2013a,Mayor2016,Ladoux2017,Hakim2017,Alert2020}. The mechanisms by which cells coordinate their motion are diverse and often not fully understood. Recent work has shown that groups of cells may respond to external stimuli as a whole, that is, in the form of collectively organized directed motion, in ways similar to what single cells do. Such collective migration can arise in response to a variety of external stimuli such as gradients in either chemical concentrations or in the stiffness of the environment, which respectively lead to collective chemotaxis \cite{Camley2016} and durotaxis.

We are interested in the phenomenon of durotaxis, which refers to the directed motion of cells along stiffness gradients of the extracellular matrix, typically towards stiffer regions. This is a well-known phenomenon for single-cell migration \cite{Lo2000}, which is rather common in many types of cells and has important implications for cancer invasion. More recently, durotaxis has been reported also for collective cell migration \cite{Sunyer2016,Sunyer2020}. Remarkably, large cell monolayers can perform durotaxis collectively even when their constituent cells do not \cite{Sunyer2016}, and in some cases, there is an optimal intermediate stiffness for tissue spreading \cite{Ng2012,Balcioglu2020b}. Collective durotaxis has been theoretically described both via hybrid computational models \cite{Escribano2018a,Gonzalez-Valverde2018,Garcia-Gonzalez2020c,Deng2020} and via a continuum active polar fluid model \cite{Alert2019a} that generalized previous work on tissue wetting \cite{Perez-Gonzalez2019}. This continuum model was solved numerically to reveal two possible mechanisms of collective durotaxis \cite{Alert2019a}.

Here, we extend the work in Ref. \cite{Alert2019a} to provide a more comprehensive classification of the dynamical regimes of the model in terms of physical parameters. Remarkably, we solve the model analytically in some simple but relevant situations, allowing for a better grasp of the physical mechanisms at play. As shown in Ref. \cite{Perez-Gonzalez2019}, the model predictions can be fitted to experimental data to infer physical parameters that are often elusive to direct measurement.

The model describes cell monolayers moving on a substrate as a quasi-twodimensional viscous fluid with two types of active forces: cell-substrate traction and cell-cell contractility. The competition between both active forces was shown to give rise to the so-called active wetting transition, whereby a tissue either spreads or retracts depending on its size \cite{Perez-Gonzalez2019}. The same model also predicted a fingering instability of the leading edge of the tissue \cite{Alert2019}. In addition to the active forces, the model also features two passive forces: an effective viscosity, which arises from cell-cell adhesion, and a friction force due to cell-substrate interactions. All these forces are treated in a coarse-grained way at the supracellular scale. The rationale of the approach is to identify the dynamical behaviours of cell monolayers that are of mechanical origin, explicitly excluding any signaling effects that cannot be encoded in the mechanical parameters of the model. To what extent such purely mechanical approach may succeed as a first step to account for the observed phenomenology is an interesting open question that might be settled by future experiments.    
\section{Hydrodynamic model}
\label{sec:2_hydro_model}

Our model stems from a hydrodynamic approach to cell tissues, a strategy that has proven useful when tissues are organized at a supracellular scale, such that information at the cellular scale is not relevant \cite{Arciero2011,Lee2011a,Lee2011,Marel2014a,Recho2016}. This is the case in many examples of collective cell migration, where coarse-grained fields such as velocity, cell density and polarization are treated as smooth fields varying on scales larger than the cell size. 
Continuum field theories based on linear irreversible thermodynamics, often called active gels theories, were first devised to account for active matter at the cellular scale, such as the cytoskeleton \cite{Kruse2005,Julicher2007,Marchetti2013,Prost2015}, but have more recently been extended to multicellular scales \cite{Julicher2018}.

The basic idea is that tissues can be modeled to some extent as continuous active materials, in such a way that the biological properties are encoded in a series of physical parameters, including passive ones such as viscosity or friction, and active ones such as contractility or traction. These parameters will in general be time and space dependent to account for the biological regulation of the cell properties and interactions. For instance, in a simple model for the spreading of epithelial monolayers \cite{Blanch-Mercader2017}, it was shown that their effective viscosity increases with time as they become thinner due to the spreading. This type of approach is useful to identify activity-driven hydrodynamic instabilities that can either be avoided or exploited by the biological regulation of parameters \cite{Alert2019,Blanch-Mercader2017c}.

\subsection{Assumptions and model equations}
\label{sec:2_model_eq}

In this paper we take the simplest possible model of an active fluid that combines active cell-substrate traction and cell-cell contractile forces. This model was introduced in Ref. \cite{Perez-Gonzalez2019} and was extended in Ref. \cite{Alert2019a} to account for substrates with non-uniform stiffness. The model is for a two-dimensional active fluid, which describes the quasi-twodimensional cell monolayer with two continuous fields: the velocity $v_\alpha$ and the polarity $p_\alpha$. The polarity is the orientational degree of freedom of the cells which arises from the polarization of its internal cystoskeletal structure and defines the direction along which traction forces are exerted. The tendency of cells to align with their neighbors is accounted for by an effective free energy of the form 
\begin{equation} \label{eq free_energy} 
    F = \int \left[\frac{a}{2}p_\alpha p_\alpha +\frac{K}{2}(\partial_\alpha p_\beta)(\partial_\alpha p_\beta) \right]\dd^2\bm{r}, 
\end{equation}
where $K$ is an effective Frank constant that quantifies the energetic cost of polarity gradients \cite{deGennes-Prost}. The constant $a$ is a restoring coefficient that is taken positive such that the unpolarized state ($p=0$) is energetically favoured in the bulk. 
We assume that the polarity follows a purely relaxational dynamics $\partial_t p_\alpha \propto - \delta F/\delta p_\alpha$ which is much faster than the temporal variations of the rest fields \cite{Perez-Gonzalez2019}, such that we can take a quasistatic evolution ($\partial_t p_\alpha = 0$) and hence $\delta F /\delta p_\alpha = 0$. Then, we have 
\begin{equation} \label{eq polarity}
    L_c^2 \nabla^2 p_\alpha = p_\alpha,
\end{equation}
where $L_c \equiv \sqrt{K/a}$ is the nematic length  that characterizes spatial variations of the polarity field \cite{Perez-Gonzalez2019,Alert2019,Alert2019a}. Since epithelial cells migrate towards free space, we enforce a boundary condition of maximum polarity $|p|=1$ directed normally to the tissue edge. Then $L_c$ defines the thickness of a polarization boundary layer near the tissue edge, such that polarity decays from $p=1$ at the edge to $p=0$ deep into the tissue.

Neglecting inertia, the force balance equation is
\begin{equation} \label{eq force_balance}
    \partial_\beta  \sigma_{\alpha\beta} + f_\alpha = 0,
\end{equation}
where $ \sigma_{\alpha\beta}$ is the stress tensor of the monolayer and $f_\alpha$ is the external force density due to the contact with the substrate. These quantities are directly related to the experimentally measured monolayer tension, $\sigma_{\alpha\beta}h$, and traction stress, $T_\alpha \equiv -f_\alpha h$, with $h$ the height of the monolayer \cite{Perez-Gonzalez2019}.

We now take the constitutive equations for a compressible active polar fluid of the form \cite{Oriola2017,Perez-Gonzalez2019}
\begin{align}
    \sigma_{\alpha\beta}&=\eta(\partial_\alpha v_\beta + \partial_\beta v_\alpha ) - \zeta p_\alpha p_\beta, \label{eq const1}\\
    f_\alpha &= -\xi v_\alpha + \zeta_i p_\alpha, \label{eq const2}
\end{align}
where $\eta$ is the viscosity, $\xi$ is the cell-substrate friction, $\zeta <0$ is the contractility, and $\zeta_i>0$ is the contact active force (hereinafter referred to as the traction parameter), which accounts for the the maximal traction stress $T_0 \equiv h\zeta_i$ exerted by polarized cells on the substrate. A summary of the symbols for the variables and parameters, together with their units and estimates, can be found in \cref{tab param}. 

We assume that, in our 2d description, the cell monolayer is compressible, with $\partial_\alpha v_\alpha \neq 0$, because the in-plane compression and expansion of the cell monolayer can be accommodated by changes in the monolayer height $h$. We assume that in-plane deformations do not amount to significant changes in pressure as the layer can deform in 3d and, hence, pressure gradients are neglected in front of the rest of the contributions in the force balance \cref{eq force_balance}. This approximation has been used and discussed for instance in Refs. \cite{Alert2019a,Perez-Gonzalez2019,Alert2019,Lee2011a,Lee2011,Blanch-Mercader2017,Blanch-Mercader2017c}. Tissue growth driven by cell proliferation is also neglected.

In the explicit form of the stress tensor \cref{eq const1} we have also assumed, following Ref. \cite{Perez-Gonzalez2019} and in order to reduce the number of parameters and define the simplest possible equations, that the bulk viscosity is $\bar{\eta}= \eta$, and that the isotropic contractility is given by $\zeta'=\zeta / 2$. Similarly, active stresses not associated to polarization are also neglected, that is, $\bar{\zeta} \ll \zeta$ as defined in Eq. (S12) in Ref. \cite{Perez-Gonzalez2019}.

Furthermore, the profile of polarity $p_\alpha$ is dictated directly by the boundary shape, so flow alignment and other elastic effects, which have been addressed in more general models such as in Ref. \cite{Blanch-Mercader2017c}, are here neglected. 
In the simplest formulation, we assume stress-free boundary conditions, but we also generalize the model to other cases.

\begin{figure}[tb!]
    \centering
    \includegraphics[width=0.91\columnwidth]{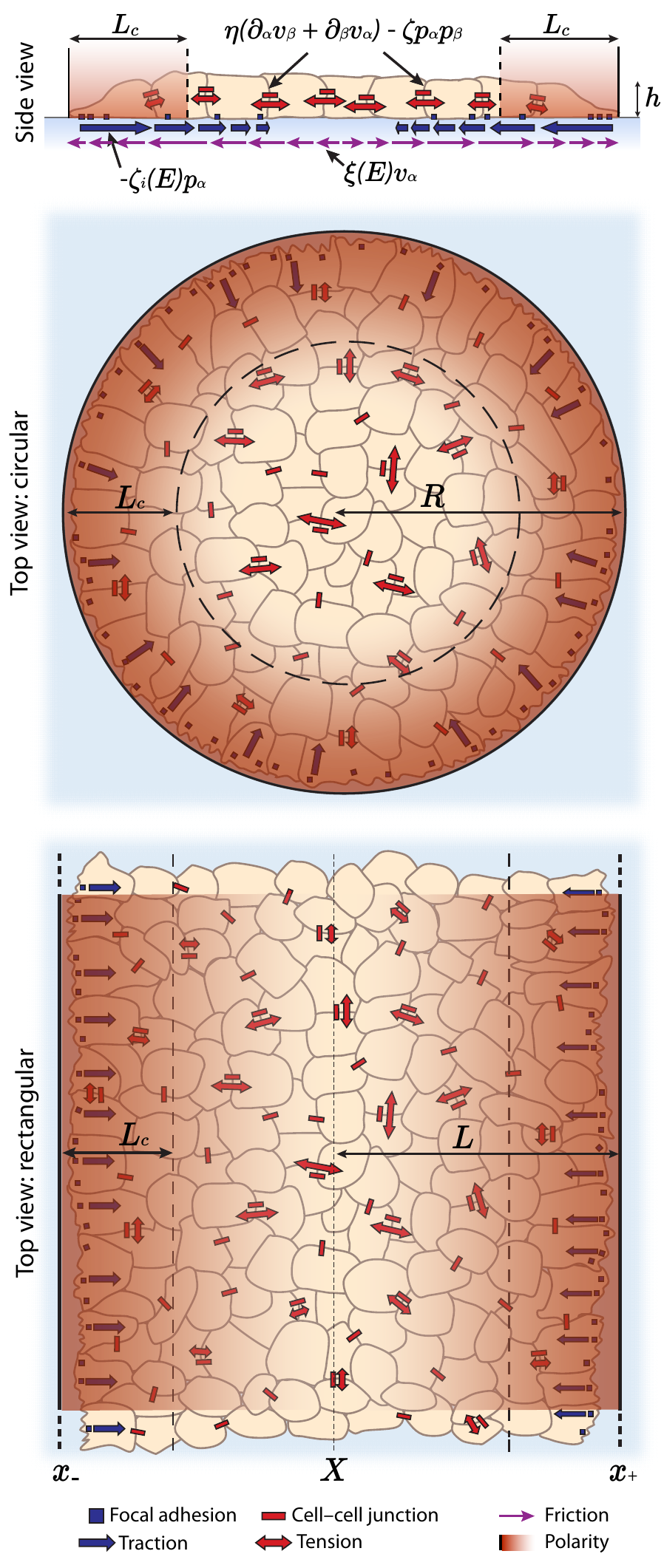}
    \caption{Scheme of the active polar fluid model for monolayer spreading, adapted from Ref. \cite{Perez-Gonzalez2019}. We model circular monolayers of radius R but reduce the description to an effective 1d setup corresponding to strips of half-width $L=R$ and infinite in the $y$ direction (see \cref{sec:2_rationale}). $X$ is the position of the center of mass of the monolayer, $x_+$ that of the right edge (stiffer when on a stiffness gradient), and $x_-$ that of the left (softer) edge. Both the traction parameter $\zeta_i$ and the friction $\xi$ (represented here being exerted on the substrate) depend on substrate stiffness, characterized by the substrate's Young modulus $E$ (see \cref{sec:3_results_durotaxis}).}
    \label{fig model_drawing}  
\end{figure}

\begin{table}[tb!] 
\begin{tabular}{lll}
 & Description [Units] & Typical value  \\
\noalign{\smallskip}\hline\noalign{\smallskip}
$\sigma_{\alpha\beta}$ & stress tensor $[ML^{-1}T^{-2}]$ &   \\ 
$f_\alpha$ & force density  $[ML^{-2}T^{-2}]$ &   \\ 
$h$  & monolayer height $[L]$  & $5$ $\mu$m \cite{Perez-Gonzalez2019,Trepat2009} \\
$L_c$ & nematic length  $[L]$ & $25$ $\mu$m \cite{Perez-Gonzalez2019,Blanch-Mercader2017} \\
$L$ & tissue half-width  $[L]$ & $200$ $\mu$m \\
$\eta$ & viscosity $[ML^{-1}T^{-1}]$& $80$ MPa·s  \cite{Perez-Gonzalez2019,Blanch-Mercader2017}\\
$\zeta$ & contractility $[ML^{-1}T^{-2}]$ & $-20$ kPa \cite{Perez-Gonzalez2019} \\
$\zeta_i^0$ & traction offset $[ML^{-2}T^{-2}]$  & $0.05$  kPa/$\mu$m \\
$\zeta_i'$ & traction gradient $[ML^{-3}T^{-2}]$ & $8\times 10^{-5} $ kPa/$\mu$m$^2$ \\
$\gamma$ & surface tension $[MT^{-2}]$  & $1$-$10$ mN/m \cite{Foty1994,Forgacs1998,Guevorkian2010,Stirbat2013,Cochet-Escartin2014,Nier2015}\\
$L_p$ & active polar length  $[L]$ & $200$ $\mu$m, $|\zeta|/(2\zeta_i)$ \\
$\lambda$ & hydrodynamic length $[L]$ & $300$ $\mu$m, $\sqrt{2\eta/\xi}$ \\
\end{tabular}
\caption{Symbols and typical values of model parameters. Here $\zeta_i^0$ and $\zeta_i'$ correspond to the linear traction profile parameters, which are adapted from the saturated profile with the stiffness and maximal traction from \cite{Sunyer2016,Alert2019a,Douezan2012c}.} 
\label{tab param}
\end{table}

\subsection{Reduction to a 1d solvable model}
\label{sec:2_rationale}

The problem at hand is formally a free-boundary problem, since the boundary of the cell cluster is free to deform and move, as its normal velocity coincides with that of the adjacent fluid. The evolution of the shape and position of the boundary is thus part of the solution of the problem. An example of how a spontaneous symmetry-breaking of the morphology of the boundary can couple to the overall motility of the domain was discussed in the context of cell fragments \cite{Blanch-Mercader2013}. In the present study we are interested in cases where the symmetry is broken by the existence of an external gradient. Since this is the dominant effect causing motion of the domain, here we ignore boundary deformations. We may also assume that the effective surface tension of the tissue is strong enough, and the monolayers small enough, to suppress the active fingering instability that is inherent to this model, as reported in Ref. \cite{Perez-Gonzalez2019,Alert2019}. Our interest is thus to describe the motion of circular monolayers of radius R on a substrate with a stiffness gradient by tracking the position of the center of mass and the monolayer size.

For simplicity, and in order to obtain exact solutions and physical insights, we formulate the problem in a 1d setup, in which the monolayers are strips that are finite in the spreading direction, and infinite in the transverse direction (\cref{fig model_drawing} bottom). This setup corresponds to the experiments on collective durotaxis of Ref. \cite{Sunyer2016} and was used also in the numerical study of the present model in Ref. \cite{Alert2019a}. The present work extends that previous study with a more comprehensive discussion of the wealth of dynamical behaviours allowed by this model and their physical interpretation, in particular taking advantage of explicit analytical solutions.

The basic physics of this 1d formulation (rectangular geometry with translational invariance in the transverse direction) is the same as that in the 2d case with circular monolayers (circular geometry with rotational invariance). The results are equivalent up to geometrical factors, but much simpler in the rectangular geometry, as already illustrated in the preceding studies in both geometries \cite{Alert2019a,Perez-Gonzalez2019,Alert2019}. Furthermore, the availability of an exactly solvable model with sufficiently simple analytical results is of great theoretical value to gain insights into the physical mechanisms at play, in particular when a relatively large number of parameters are present. Moreover, we will show that some of the limitations of the 1d formulation, such as the lack of the Young-Laplace pressure drop due to tissue surface tension, can be effectively introduced in a simple way into the 1d reduction of the problem.   

In the 1d setup, \crefrange{eq force_balance}{eq const2} reduce to
\begin{equation}
2\eta \partial_x^2 v = 2 \zeta p \partial_x p + \xi v - \zeta_i p.  \label{eq main}
\end{equation}
The polarity profile is given by the solution of \cref{eq polarity} satisfying $p=\pm 1$ at the the two edges $x=x_+$ and $x=x_- < x_+$ respectively. In terms of the center of mass position $X \equiv (x_+ +x_-)/2$ and the monolayer half-width $L \equiv (x_+ - x_-)/2$, it reads
\begin{equation} \label{eq polarity_sol}
p(x)= \frac{\sinh{((x-X)/L_c)}}{\sinh{(L/L_c)}}.
\end{equation}

There are several length scales whose ratios determine different physical scenarios in the model. The scale $L_c$ is typically the smallest one, as the polarized boundary layer of the tissue is often thin compared to the system size $L$ and the other length scales \cite{Sunyer2016,Perez-Gonzalez2019,Blanch-Mercader2017}. The so-called screening length $\lambda \equiv \sqrt{2\eta/\xi}$ is a measure of the range of hydrodynamic interactions \cite{Marchetti2013,Blanch-Mercader2017,Alert2019,Alert2019a}, and it defines two important limits: In the so-called `wet' limit, when $\lambda \gg L$, long-ranged hydrodynamic interactions produce non-local effects and the system behaves globally as a whole; in the so-called `dry' limit, when $\lambda \ll L$, the spreading dynamics is governed by local forces, \textit{i.e.} the two edges behave independently from each other. Another relevant length scale, that we call the active polar length $L_p$ \cite{Perez-Gonzalez2019}, arises as the ratio of contractility to traction forces: $L_p \equiv |\zeta|/ (2\zeta_i)$. In the wet case, this length defines the critical tissue size for the wetting-dewetting transition, as reported in Ref. \cite{Perez-Gonzalez2019}.

\Cref{eq main} will be solved typically with stress-free boundary conditions. If a normal stress component is required to mimic the effect of an effective surface tension, as if $L$ would be the monolayer radius, we will impose
$\sigma_{\pm} = -\gamma/L$, which implies $\partial_x v |_{x_{\pm}}= (\zeta - \gamma/L )/(2\eta)$. The solution of \cref{eq main} provides the spatial velocity profile $v(x)$, from which we obtain $v_{\pm}$ as well as the velocity of the center of mass $U\equiv\dot{X}$ and the spreading velocity $V\equiv\dot{L}$.

For tissues on a substrate with variable stiffness, the parameters of the passive and active forces on the surface, that is friction and active traction, will be space-dependent, $\xi(x)$ and $\zeta_i(x)$. The relationship between these spatial variations and that of the substrate stiffness must be determined independently of the hydrodynamic model. An explicit derivation requires a detailed knowledge of the molecular mechanisms at play, and a discussion based on empirical data was made for instance in Ref. \cite{Alert2019a}. Both friction and traction parameters increase with and eventually saturate with increasing substrate stiffness \cite{Walcott2010,Saez2010,Trichet2012,Gupta2015,Marcq2011,Sens2013a}. However, to avoid introducing more parameters and to make the interpretation of the results more transparent, we mostly consider cases where those parameters are either space-independent or have a uniform gradient, hence introducing only two new parameters associated to the stiffness variation, namely  
$\xi'\equiv \partial_x \xi(x)$ and $\zeta_i'\equiv \partial_x \zeta_i(x)$. This restriction is relaxed in \cref{sec:3_results_durotaxis_friction}.
In most cases we take $\xi'=0$ (uniform friction) and focus on the effect of a uniform traction gradient $\zeta_i'>0$ on the net displacement of the monolayer. We then find the velocity profile at any given time by solving \cref{eq main} with the initial conditions $L_0\equiv L(0)$ and $X_0\equiv X(0)$ for the set of parameters $L_c, \eta, \xi, \zeta, \zeta_i^0, \zeta_i'$, where $\zeta_i^0 \equiv \zeta_i(X_0)$ is the initial traction offset.

\subsection{Solutions for a uniform substrate}
\label{sec:2_results_uniform}

We first consider as a reference the case with no stiffness gradient, so that $\zeta_i'=0$,  $\xi'=0$, and consequently there is no net monolayer displacement: $U=0$. This case was studied in the wet limit, $\xi\rightarrow 0$ in Ref. \cite{Perez-Gonzalez2019} in circular geometry, and in the wet-dry crossover and in rectangular geometry in Ref. \cite{Alert2019}. The exact solution of this case is given \cref{app const_substrate}. Taking $\gamma=0$ and assuming $L_c \ll L$, the expression in the wet limit $\lambda \gg L$ for the spreading velocity takes the simple form
\begin{equation} \label{eq const_wet}
V^{wet} = \pm v_{\pm}^{wet} \approx  \frac{L_c}{2\eta} \left[ L\zeta_i - \frac{|\zeta|}{2}\right] = 
\frac{L_c\zeta_i}{2\eta} \left(L - L_p\right),
\end{equation}
which recovers Eqs. (5) and (7) from Ref. \cite{Alert2019}. In the dry limit $L_c \ll \lambda \ll L$, we obtain
\begin{equation} \label{eq const_dry}
V^{dry} = \pm v_{\pm}^{dry} \approx  \frac{L_c}{2\eta} \left[ \lambda \zeta_i - \frac{|\zeta|}{2}\right] = 
\frac{L_c\zeta_i}{2\eta} \left( \lambda - L_p\right).
\end{equation}

\begin{figure}[tb!]
  \centering
  \includegraphics[width=0.85\columnwidth]{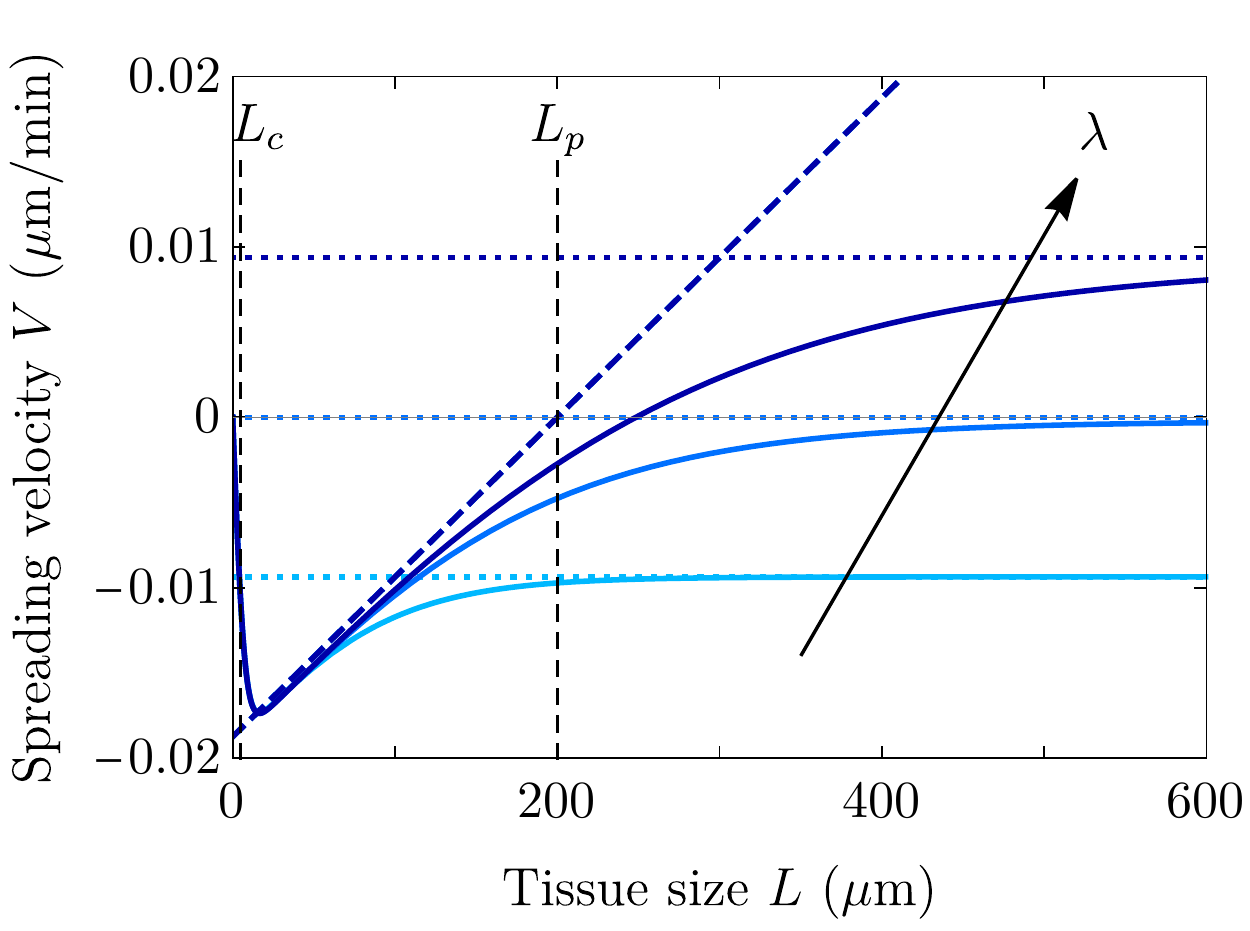}
  \caption{Spreading velocity as a function of the monolayer half-width on a uniform substrate, for $L_p = 200$ $\mu$m and $\lambda=100,200$ and $300$ $\mu$m. Solid lines show the full expressions in \cref{eq const_full}, and the dotted and the dashed lines are the dry and the wet limits, respectively, which converge to the full expressions at large sizes. Parameter values are in \cref{tab param}, except for $L_c = 5$ $\mu$m (smaller to see better convergence). Only for the largest $\lambda$, the critical size $L^* \approx 200 \mu\text{m}=L_p$ approaches the wet limit prediction; for the other two values of $\lambda$, the dry approximation is better for smaller sizes.}
  \label{fig vsL0_lambda}
\end{figure}

In the wet limit, there is a critical tissue size $L^*\approx L_p$ that defines the so-called active wetting transition of Ref. \cite{Perez-Gonzalez2019}. This transition distinguishes whether the cluster is expanding (positive spreading velocity $V>0$ for $L>L^*$) or contracting (negative spreading velocity $V<0$ for $L<L^*$). The wet limit is represented by the dashed line in \cref{fig vsL0_lambda}. In this limit, the spreading velocity $V$ does not depend on $\lambda$, and the transition from contraction to expansion takes place at $L= L_p$. The condition $V=0$ thus defines the wetting-dewetting transition reported in Ref. \cite{Perez-Gonzalez2019}. However, here we refer to spreading and avoid the term `wetting', which usually refers to the local motion of a fluid front on a substrate. This precision is meant to avoid confusion in cases where the center of mass of the tissue is moving. In those cases, the soft tissue edge may recede with respect to the substrate while the tissue globally expands. We discuss such examples in \cref{sec:3_results_durotaxis_linear}.

In the dry limit, the spreading transition is controlled by the screening length $\lambda$. \Cref{eq const_dry} shows that there is a critical $\lambda^* \approx L_p$ such that for $\lambda < \lambda^*$ the cluster contracts ($V<0$), regardless of its size $L$, and for $\lambda > \lambda^*$ the cluster always expands ($V>0$). This result in the dry limit is represented by the dotted lines in \cref{fig vsL0_lambda}, which do not depend on the tissue size $L$ and exhibit the spreading transition at $V = 0$ for $\lambda=L_p$.

\begin{figure}[tb!]  
  \includegraphics[width=.50\columnwidth, trim={0 0cm 0 0},clip]{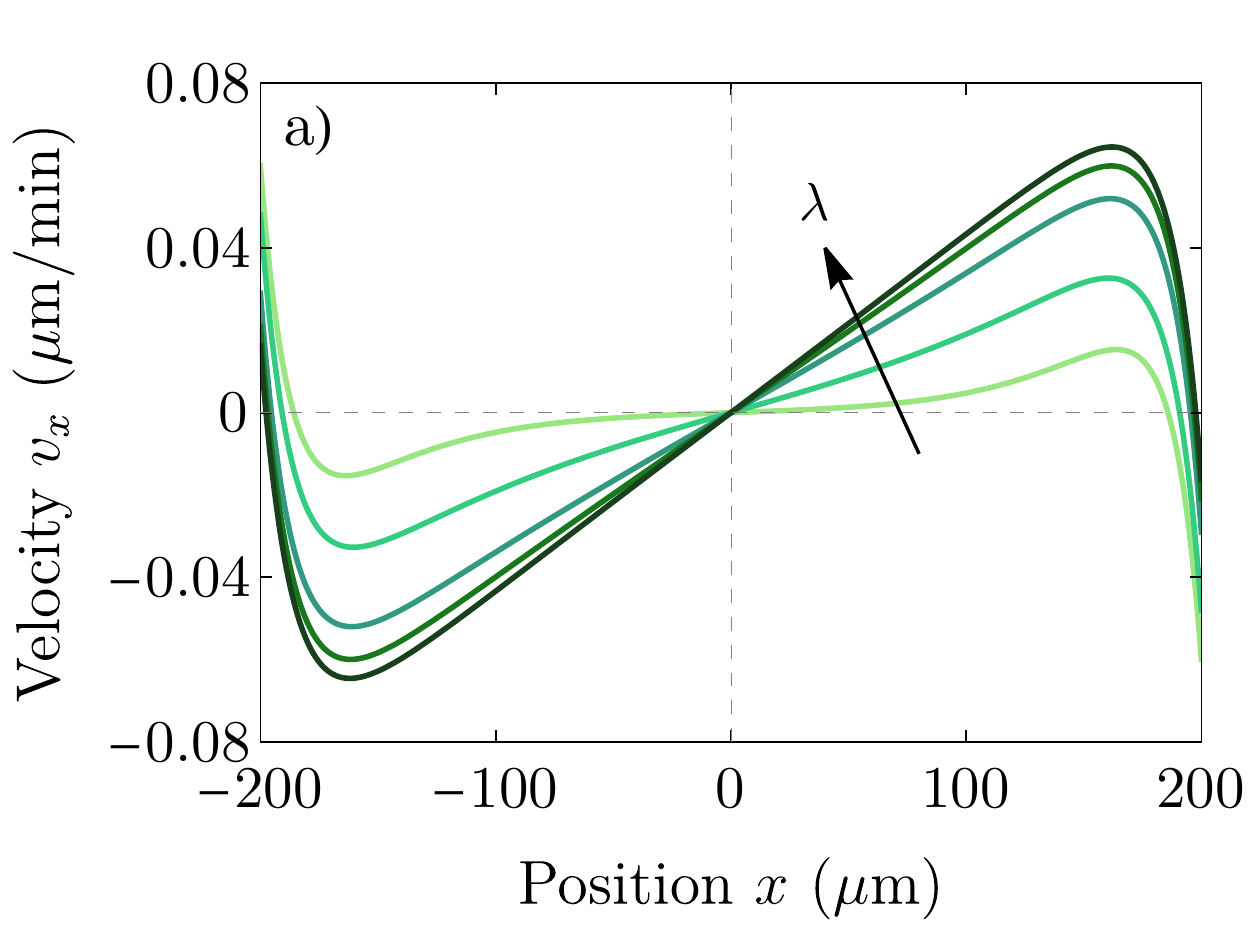}\hfill
  \includegraphics[width=.48\columnwidth, trim={0 0 0 0},clip]{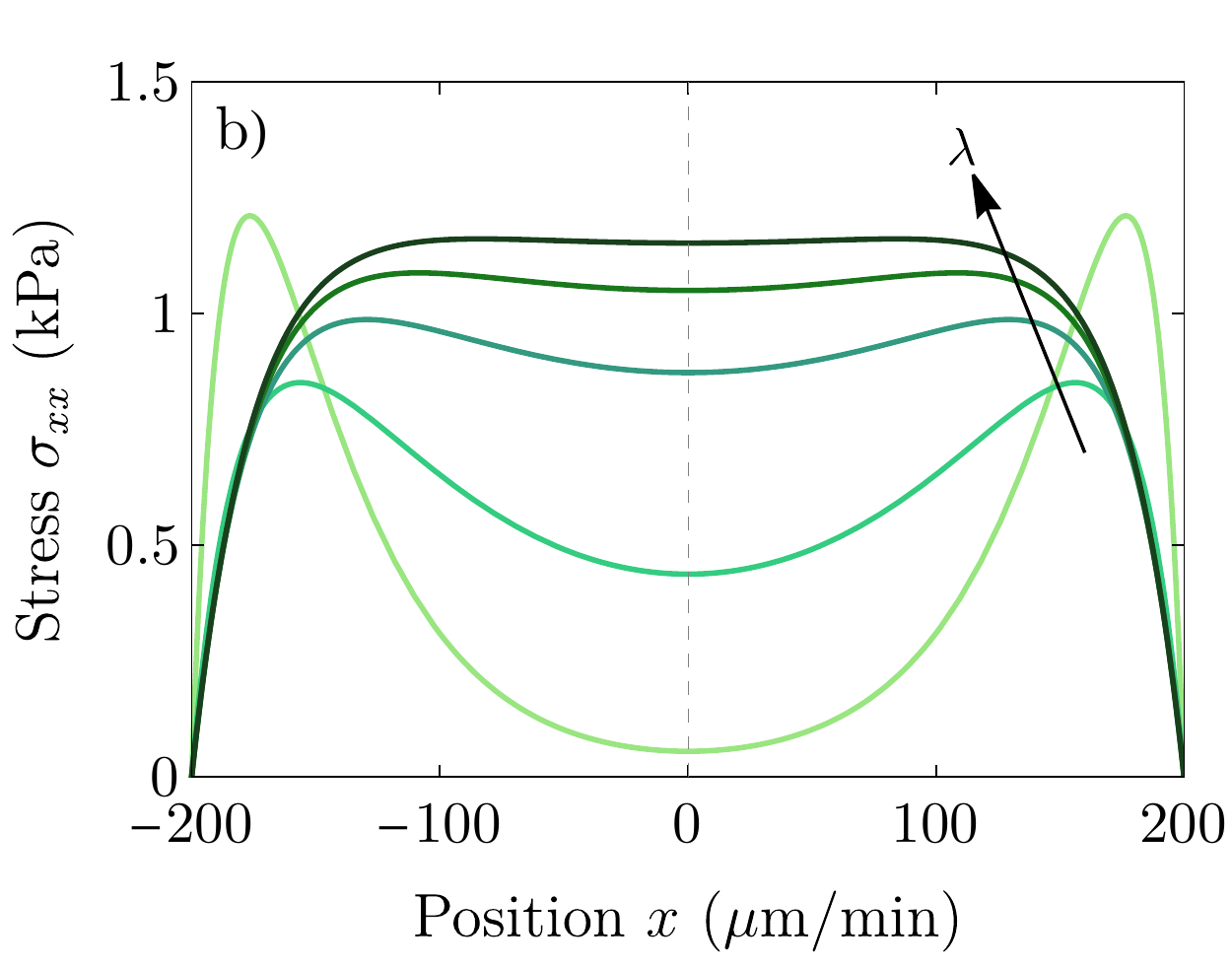}\hfill
  \caption{Velocity (a) and stress (b) profiles for the case of a uniform-stiffness substrate. Parameter values are given in \cref{tab param} except for $\lambda$, which takes values $\lambda = 40,100,200,300,450$ $\mu$m.} 
  \label{fig velstress_profile_const}
\end{figure} 

The full velocity and stress profiles (\cref{fig velstress_profile_const}) are not used here, but they allow the model predictions to be tested against experimental data. These profiles provide a simple visualization of where the system stands in the wet-dry axis, and of the forces and flows in the cell monolayer. For example, the stress plateau in the bulk (darkest curves in \cref{fig velstress_profile_const}b) is a signature of the wet limit (large $\lambda$). Respectively, two peaks of width $L_c$ near the edges (lighter curves in \cref{fig velstress_profile_const}b) are indicative of the dry limit (small $\lambda$). In this case, the velocity profiles features a plateau of null velocity in the bulk (lightest curve in \cref{fig velstress_profile_const}a). In this situation with a uniform substrate stiffness, the profiles of stress and velocity are respectively even and odd with respect to the center of the monolayer.

\section{Collective durotaxis}
\label{sec:3_results_durotaxis}

\subsection{Linear traction profile}
\label{sec:3_results_durotaxis_linear}

The presence of a stiffness gradient should in general affect the interactions between the cells and the substrate, thus altering both traction and friction forces. The dependence of the traction and friction parameters on substrate stiffness must be determined independently of the model, either empirically or from a microscopic model of cell-substrate interactions. In this section, to obtain analytical solutions, we take the simplest possible spatial dependence of these parameters: a linear profile of traction $\zeta_i(x) = \zeta_i^0 + \zeta_i' (x-X) $, and a uniform friction coefficient, with $\xi'=0$. The corresponding results will be applicable locally to more general traction profiles as long as $\zeta_i''L/\zeta_i' \ll 1$. The exact results for a linear traction profile, together with some approximate expressions, are given in \cref{app grad_substrate}.

An important exact result for this case of uniform traction gradient $\zeta_i'$ is that the spreading velocity $V$ is that on a uniform substrate $V^u$ with the traction evaluated at the monolayer center, that is $V(\zeta_i^0,\zeta_i')=V^u(\zeta_i^0)$. Therefore, the spreading behavior is independent of the existence of a traction gradient. More generally, in cases where the traction gradient is not quite uniform, the spreading velocity will be relatively insensitive to that gradient. The velocity of the center of mass, however, is sensitive to the existence of a traction gradient, which gives rise to the phenomenon of durotaxis. 

\begin{figure}[tb!]
  \centering
  \includegraphics[width=0.85\columnwidth]{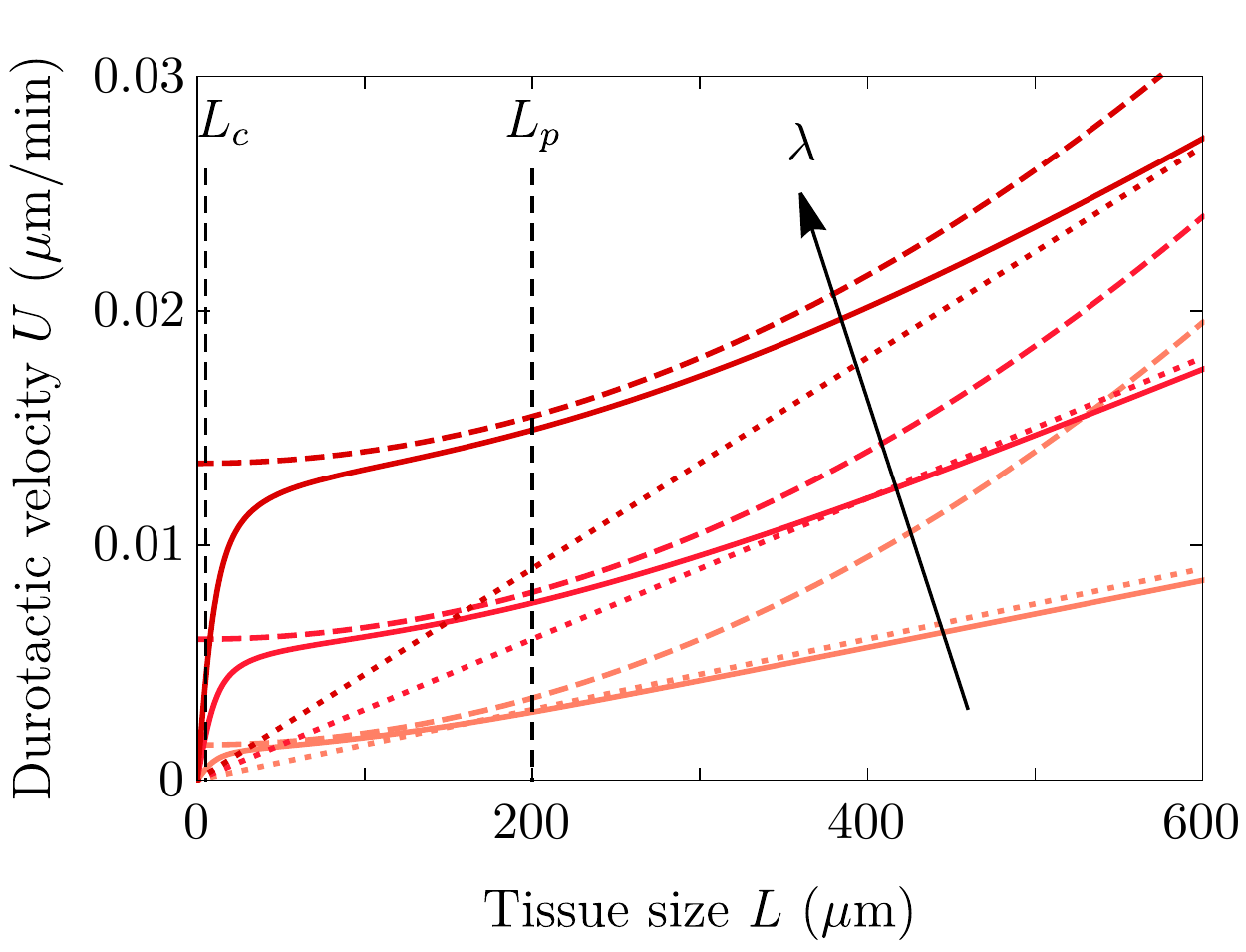}
  \caption{Center-of-mass velocity as a function of the monolayer half-width for a linear traction profile, with $L_p = 200$ $\mu$m and $\lambda=100,200$ and $300$ $\mu$m. As in \cref{fig vsL0_lambda}, the solid lines represent the full expression in \cref{eq grad_full_v0}, and the dotted and dashed lines represent the dry and the wet limits, respectively. Parameter values are in \cref{tab param}, except for $L_c = 5$ $\mu$m.}
  \label{fig v0L0_lambda}
\end{figure}

Next, we discuss the results in the dry and wet limits. Taking $\gamma=0$, the expression of the edge velocities in the dry limit  ($L_c \ll \lambda \ll L$) reads
\begin{align}  \label{eq grad_dry}
    v_\pm^{dry}  &\approx \pm \frac{L_c}{2\eta}\left[\lambda \zeta_i^{\pm} - \frac{|\zeta|}{2} \pm 2\zeta_i'L_c^2 \right] \nonumber \\
    &= v_{\pm}^{u,dry}(\zeta_i^{\pm}) + \frac{\zeta_i'L_c^3}{\eta}
\end{align} 
where $\zeta_i^{\pm}$ are the local values of the traction at the edges. The corresponding center-of-mass velocity, neglecting $2L_c^2$ in front of $L\lambda$, reads 
\begin{equation} \label{eq grad_dryv0}
    U^{dry}  \approx  \frac{L_c\lambda}{2\eta} L\zeta_i' = \frac{L_c \lambda }{4\eta} (\zeta_i^+-\zeta_i^-), 
\end{equation}
and the spreading velocity is $V^{dry} = V^{u,dry}(\zeta_i^0)$,
with $\zeta_i^0 = (\zeta_i^++\zeta_i^-)/2$. 
Although $L$ appears in \cref{eq grad_dryv0}, giving a linear increase of $U$ with $L$ (dotted lines in \cref{fig v0L0_lambda}), $U$ can be rewritten in terms of the traction difference emphasizing that the spreading dynamics is local in the sense that the two edges behave independently from the other. The traction difference then directly drives tissue durotaxis. 

On the contrary, in the wet limit\footnote{The strict wet limit $\lambda\rightarrow \infty$ ($\xi \rightarrow 0$) with finite $L_c$ is ill-defined for a the case of a nonzero $\zeta_i'$, because force balance cannot be globally satisfied unless there is friction.} $L_c \ll L \ll \lambda$, the two edges are coupled through hydrodynamic interactions, and we get 
\begin{align}  \label{eq grad_wet}
    v_{\pm}^{wet}  &\approx \pm \frac{L_c}{2\eta}\left[L \zeta_i^{\pm}  - \frac{|\zeta|}{2} \pm \zeta_i'\Big( \lambda^2 - \frac{2}{3}L^2 \Big) \right] \nonumber \\ &= v_{\pm}^{u,wet}(\zeta_i^{\pm}) + \frac{L_c \zeta_i'}{2\eta}\Big( \lambda^2 - \frac{2}{3}L^2 \Big),  
\end{align}
which yields a center-of-mass velocity
\begin{equation}  \label{eq grad_wetv0}
    U^{wet}  \approx  \frac{\zeta_i'L_c}{2\eta} \Big( \lambda^2 +\frac{ L^2}{3}\Big) \approx \frac{L_c \lambda }{2\eta} \lambda \zeta_i', 
\end{equation}
and a spreading velocity $V^{wet} =  V^{u,wet}(\zeta_i^0)$. Both $v_\pm$ and $U$ depend on the system size $L$ and the traction gradient $\zeta_i'$, which illustrates that the two edges are hydrodynamically coupled. We provide a summary of results for tissue durotaxis $U$ and spreading $V$ in the wet and dry limits in \cref{tab results}.

Two main conclusions emerge which are general in the whole wet/dry range for this case of uniform traction gradient $\zeta_i'$ and uniform friction ($\xi'=0$). On the one hand, the center-of-mass velocity $U$ is proportional to the traction gradient $\zeta_i'$ and independent of the traction offset $\zeta_i^0$. $U$ has the same sign as $\zeta_i'$, and there is durotaxis to stiffer regions as long as the traction is a monotonically increasing function of the stiffness. On the other hand, the spreading velocity depends on the traction offset and not on the traction gradient. Accordingly, \cref{fig vsL0_lambda} still applies in the present case, and durotaxis is independent of whether the monolayer is spreading or contracting (\cref{fig v0L0_lambda}).

In fact, the following situations are possible. First, the monolayer can contract either with the two edges moving in opposite directions ($v_->0$ and $v_+<0$) or in the same direction ($0 < v_+ < v_-$). In the former case, both edges are retracting, or dewetting. In the latter case, the $+$ edge is wetting and the $-$ edge is dewetting. Second, the monolayer can expand, or spread, if both edges move away from each other ($v_-<0 $ and $v_+ > 0$), both wetting the substrate, but also if both edges move in the same direction ($0 < v_- < v_+$), with the $+$ edge wetting and the $-$ one dewetting.

\begin{table}[tb!]
\centering
\begin{tabular}{ccccc}
&  & \multicolumn{1}{c}{Uniform} & \multicolumn{2}{c}{Stiffness gradient} \\
&  & \multicolumn{1}{c}{\small{($\zeta_i,\xi$ unif.)}} & \multicolumn{2}{c}{\small{(lin. $\zeta_i(x)$, unif. $\xi$)}} \\
\hline 
\rule{0pt}{3ex}    
\multirow{2}{*}{Wet ($L\ll \lambda$)} 
    & \multicolumn{1}{c|}{$U$} & \multicolumn{1}{c}{0} & \multicolumn{2}{c}{$\frac{L_c\lambda}{2\eta}\lambda \zeta_i'$} \\ \cline{2-5} 
    \rule{0pt}{3ex}    
    & \multicolumn{1}{c|}{$V$} & \multicolumn{1}{c}{$\frac{L_c\zeta_i}{2\eta}(L-L_p)$} & \multicolumn{2}{c}{$\frac{L_c\zeta_i^0}{2\eta}(L-L_p)$}   \\
\hline
\rule{0pt}{3ex}    
\multirow{2}{*}{Dry ($\lambda \ll L$)} 
    & \multicolumn{1}{c|}{$U$} & \multicolumn{1}{c}{0} & \multicolumn{2}{c}{$\frac{L_c\lambda}{2\eta}L \zeta_i'$ } \\ \cline{2-5} 
    \rule{0pt}{3ex}    
    & \multicolumn{1}{c|}{$V$} & \multicolumn{1}{c}{$\frac{L_c\zeta_i}{2\eta}(\lambda-L_p)$} & \multicolumn{2}{c}{$\frac{L_c\zeta_i^0}{2\eta}(\lambda-L_p)$} \\
\hline
\end{tabular}
\caption{Summary of the results. In the wet limit, the critical length is such that if $L<L^* \approx L_p $ there is contraction ($V<0$) whereas if $L>L^*$ there is expansion ($V>0$). In the dry limit both regimes are defined by $\lambda^* \approx L_p$.}
\label{tab results}
\end{table}

\begin{figure}[tb!]
  \centering
    \includegraphics[width=0.85\columnwidth, trim={0cm 1cm 0 0},clip]{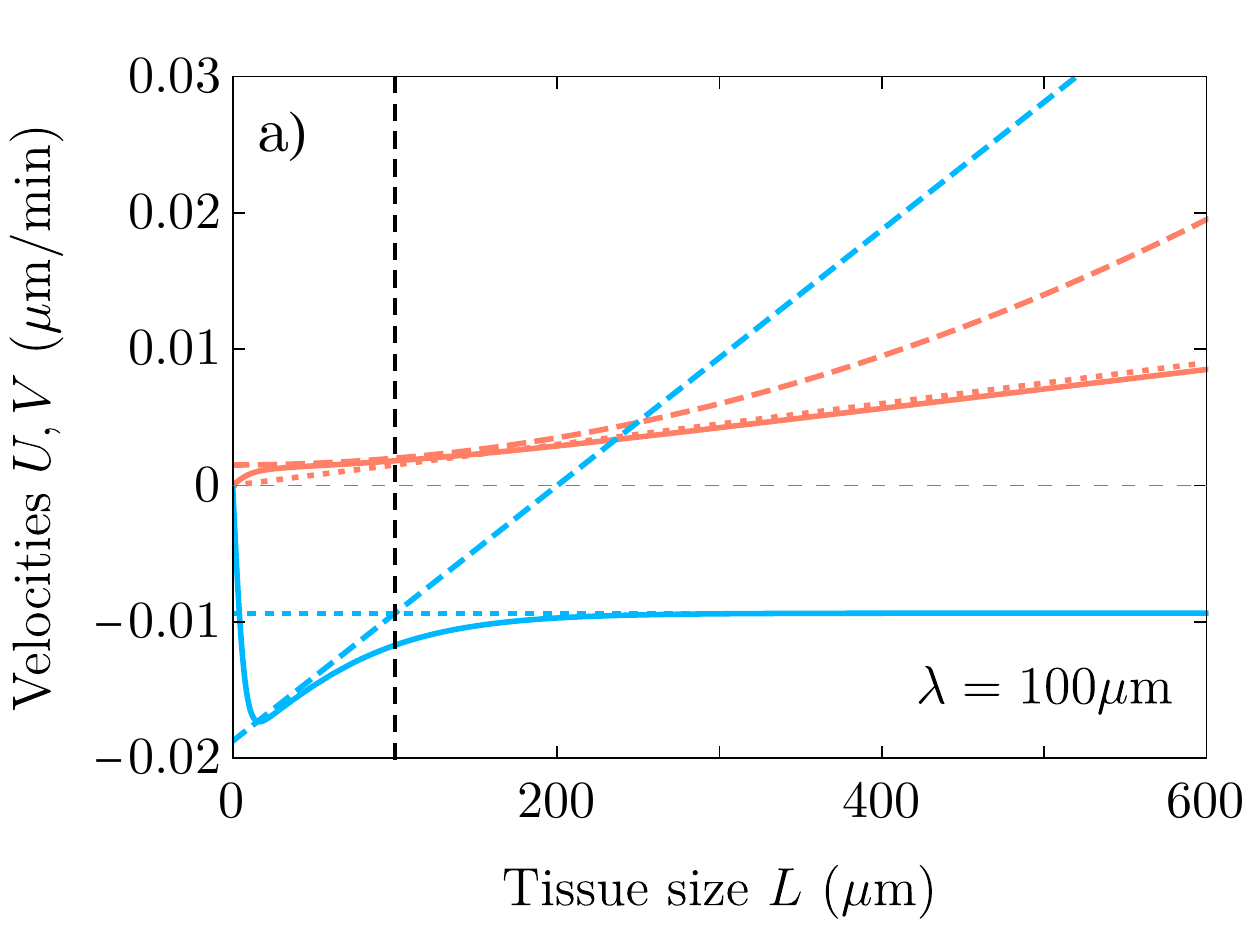} \vfill
  \includegraphics[width=0.85\columnwidth, trim={0cm 1cm 0 0},clip]{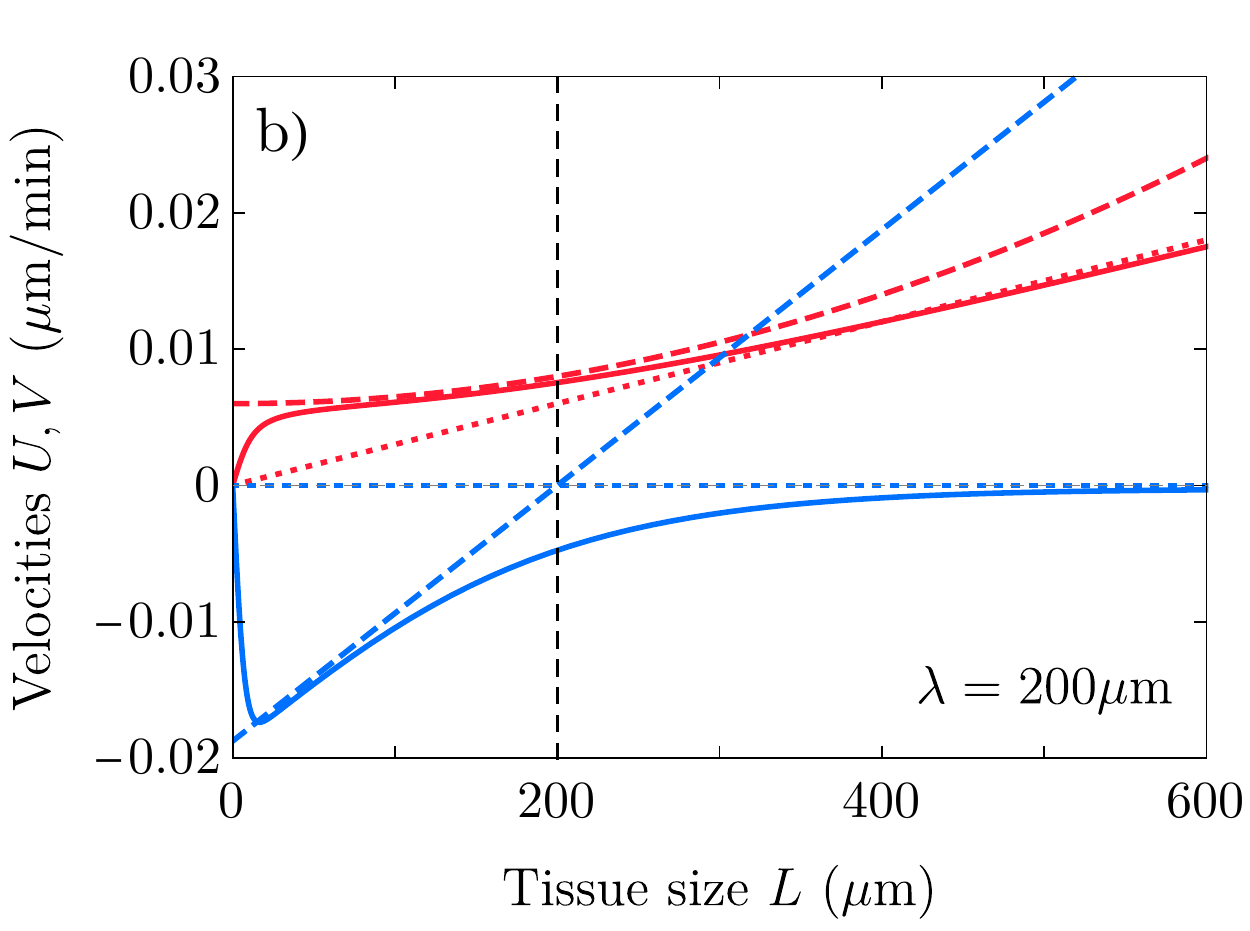} \vfill
  \includegraphics[width=0.85\columnwidth]{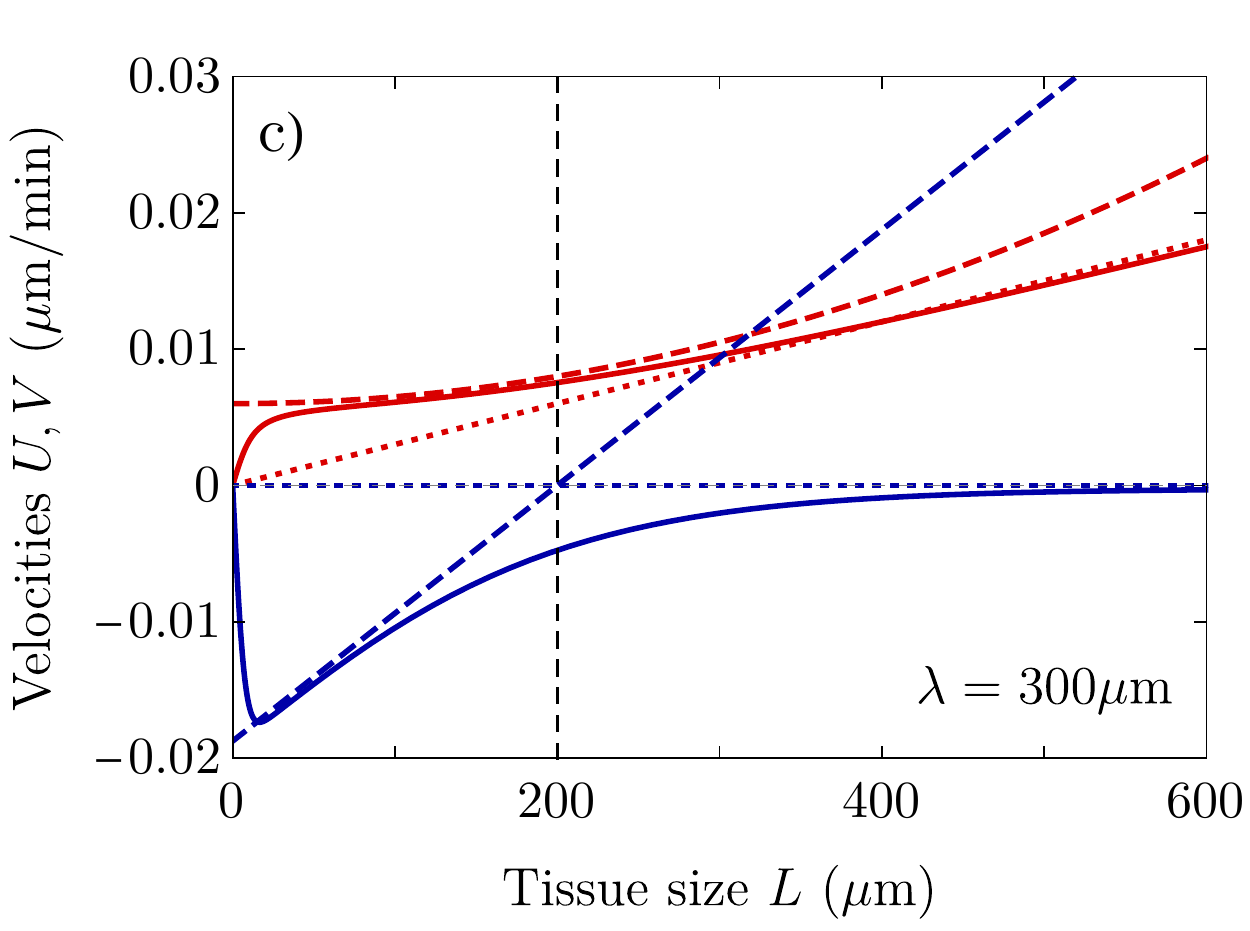}
  \caption{Spreading velocity $V$ (blue) and center-of-mass velocity $U$ (red) in their full expressions (solid), dry (dotted) and wet (dashed) limits, for three different values of $\lambda$ (vertical dashed lines) and for $L_p=200$ $\mu$m. Each figure corresponds to the $V$ and $U$ in \cref{fig vsL0_lambda,fig v0L0_lambda} for its particular $\lambda$. Parameter values are in \cref{tab param}, except for $L_c = 5$ $\mu$m. Together with \cref{fig v0vs_lambdacases_edges}, these plots show that the monolayer contracts with both edges dewetting for all $L$ in (a), for $L \lessapprox 127$ $\mu$m in (b) and for $L \lessapprox 54$ $\mu$m in (c) (solutions of $v_+=0$ in the wet predictions). It contracts with the $+$ edge wetting but the $-$ edge dewetting faster for $L \gtrapprox  127$ $\mu$m in (b) and for $ 54$ $\mu$m $\lessapprox L \lessapprox 200$ $\mu$m in (c) (solutions of $v_+=v_-$ in the wet predictions). Finally, the monolayer expands with the $-$ edge dewetting but slower than the $+$ one wets for $L \gtrapprox 200$ $\mu$m in (c). To have an expanding monolayer with both edges wetting the substrate, we should set lower contractilities or larger tractions.}
  \label{fig v0vs_lambdacases}
\end{figure}

It is thus clear that the condition of spreading or contraction, which is a property of the cell monolayer as a whole, and the condition of wetting or dewetting, which refers to the direction of motion of each tissue edge, are two distinct conditions that only coincide when the center of mass does not move ($U=0$), as in Ref. \cite{Perez-Gonzalez2019}. We show examples of these distinct situations in \cref{fig v0vs_lambdacases} and \cref{fig v0vs_lambdacases_edges} in \cref{app v0vs_edges}.

\begin{figure}[tb!]
  \includegraphics[width=.52\columnwidth]{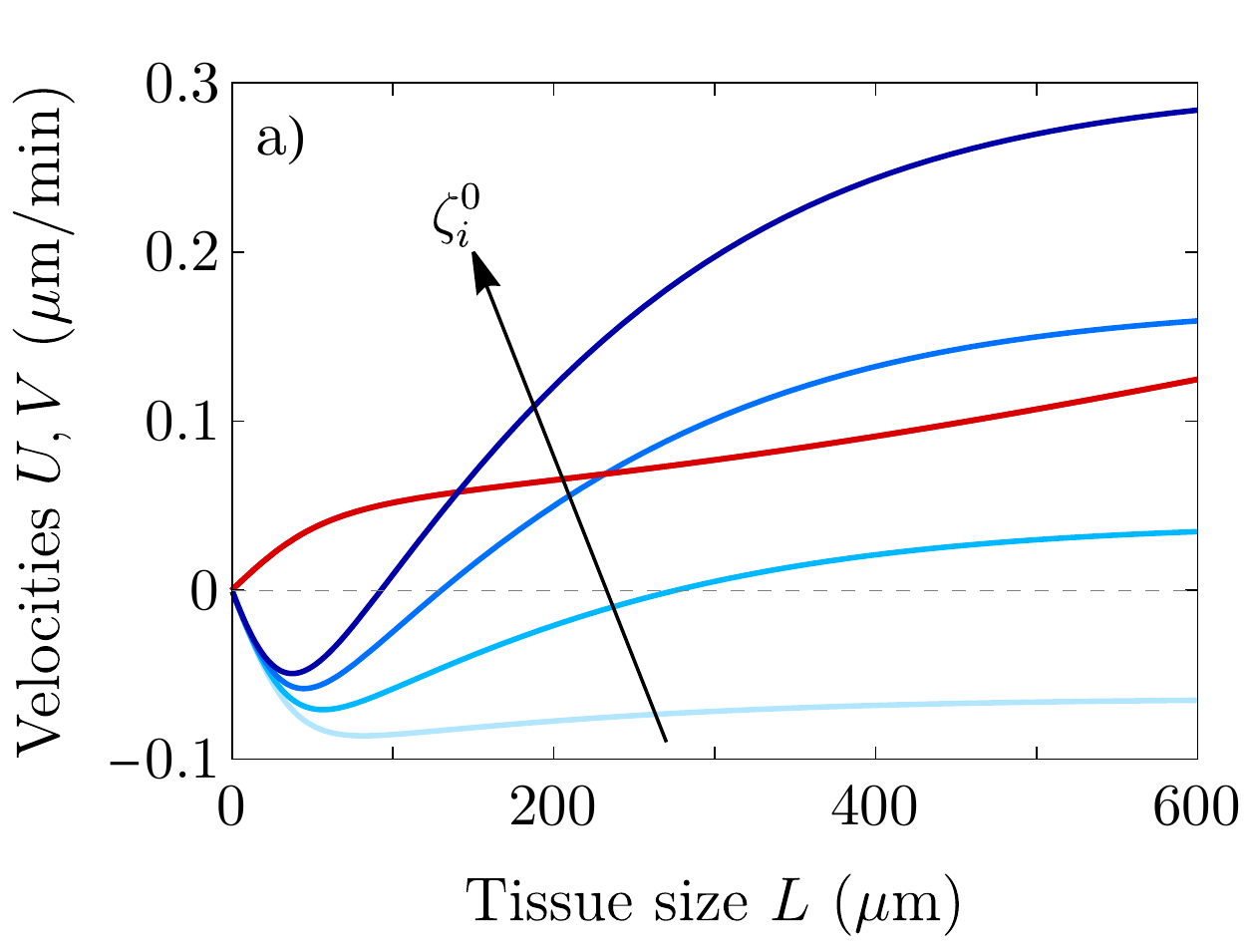}\hfill
  \includegraphics[width=.48\columnwidth,trim={0.8cm 0 0 0},clip]{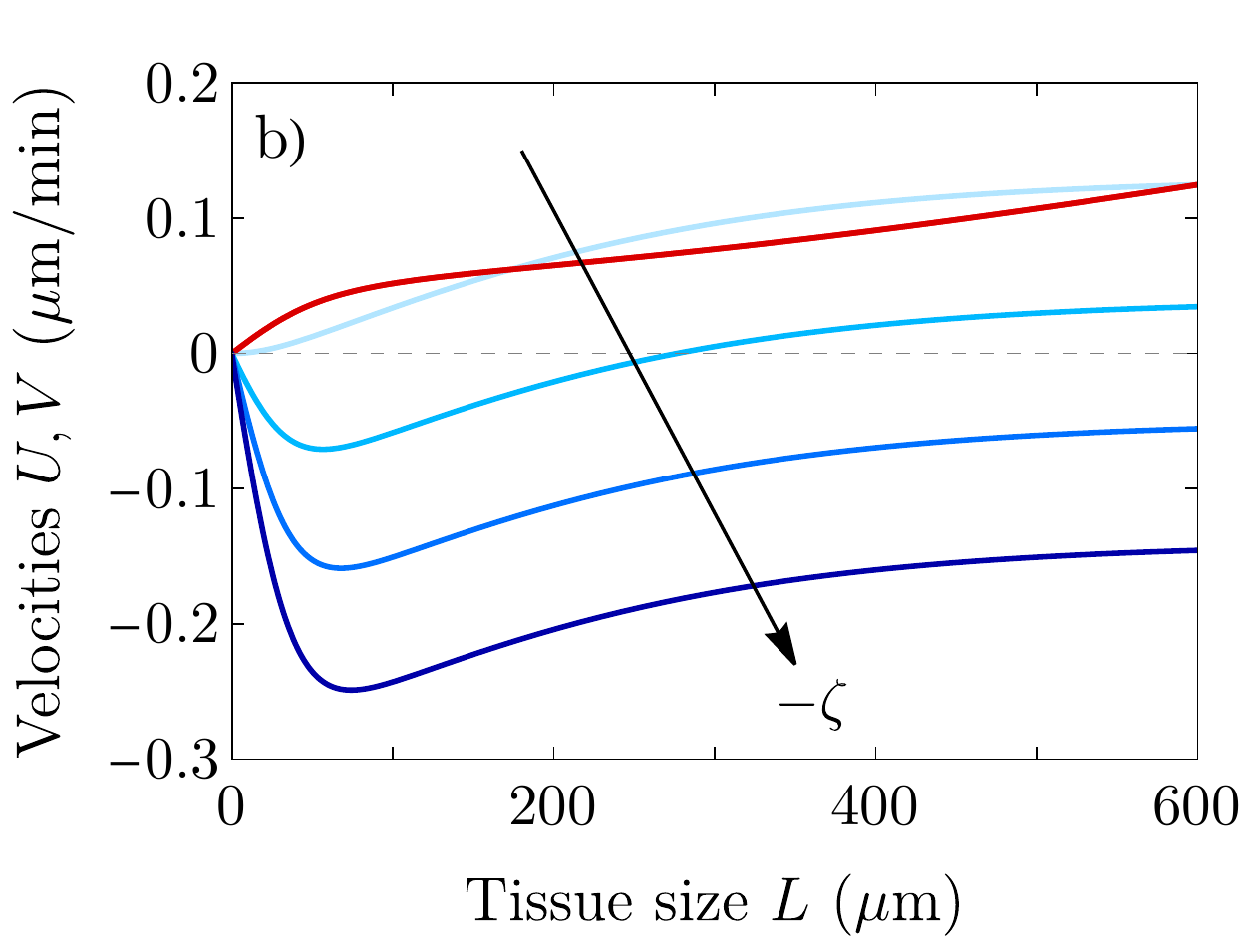}
  \caption{Plots of the full expressions for $V$ (blue curves) and $U$ (red curves) as a function of tissue size $L$ for the parameter values in \cref{tab param} but changing (a) the traction offset $\zeta_i^0 = 0.01, 0.05, 0.10, 0.15$ kPa/$\mu$m, and (b) the contractility $-\zeta = 0, 20, 40, 60 $ kPa.}
  \label{fig vvsL0_param}
\end{figure}

\begin{figure}[tb!]
  \includegraphics[width=.515\columnwidth, trim={0 0.9cm 0 0},clip]{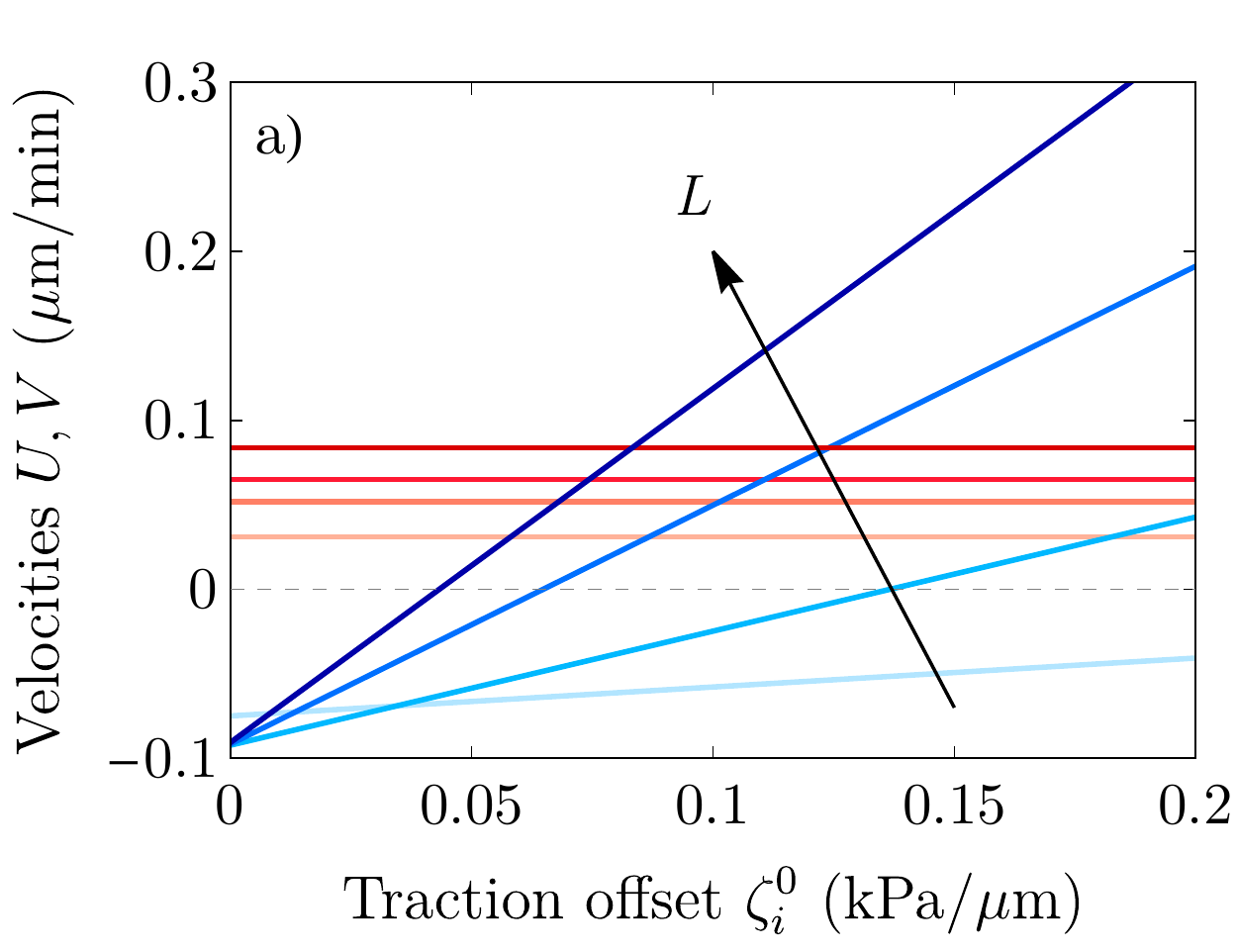}\hfill
  \includegraphics[width=.485\columnwidth, trim={0.8cm 0.9cm 0 0},clip]{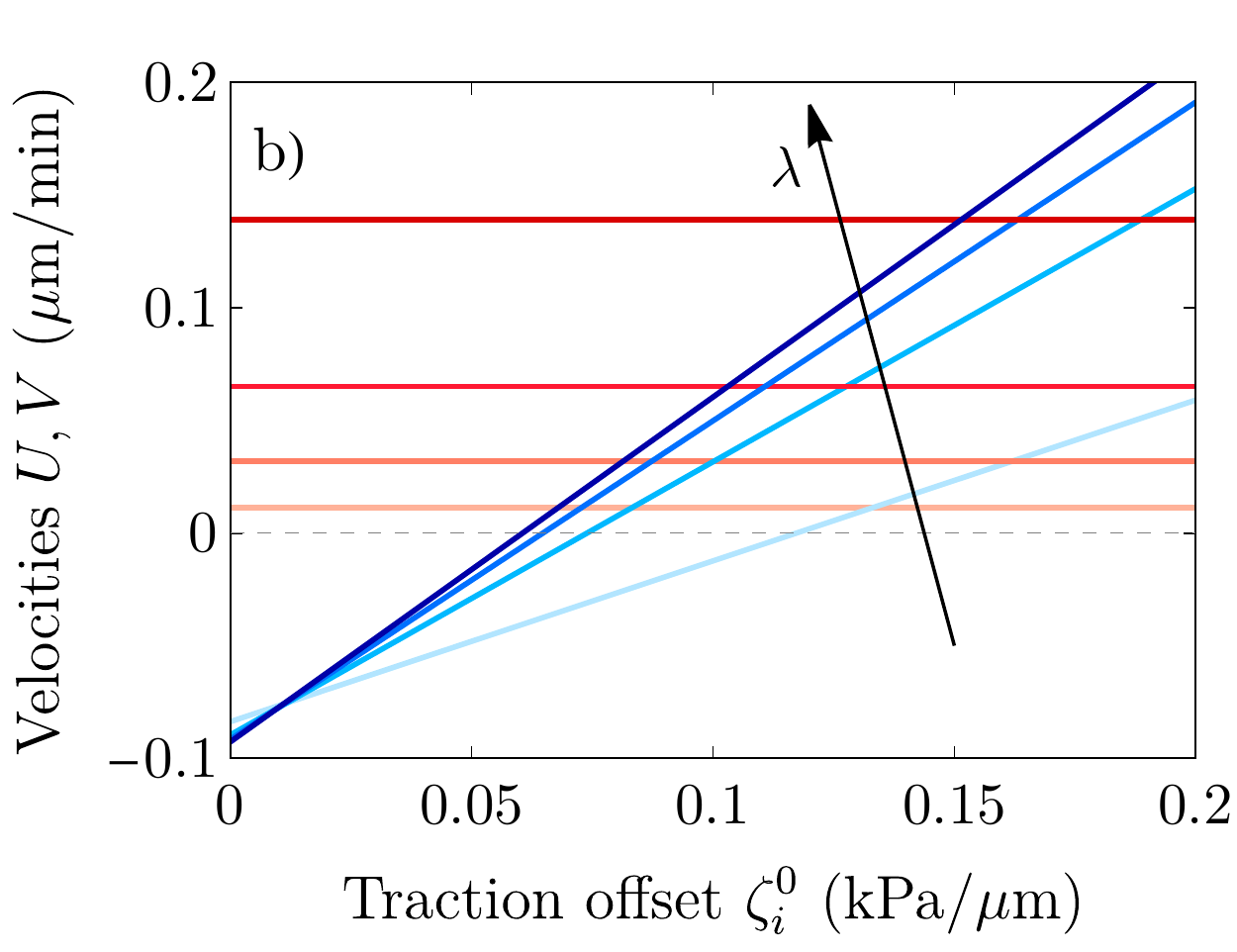}
  \includegraphics[width=.515\columnwidth]{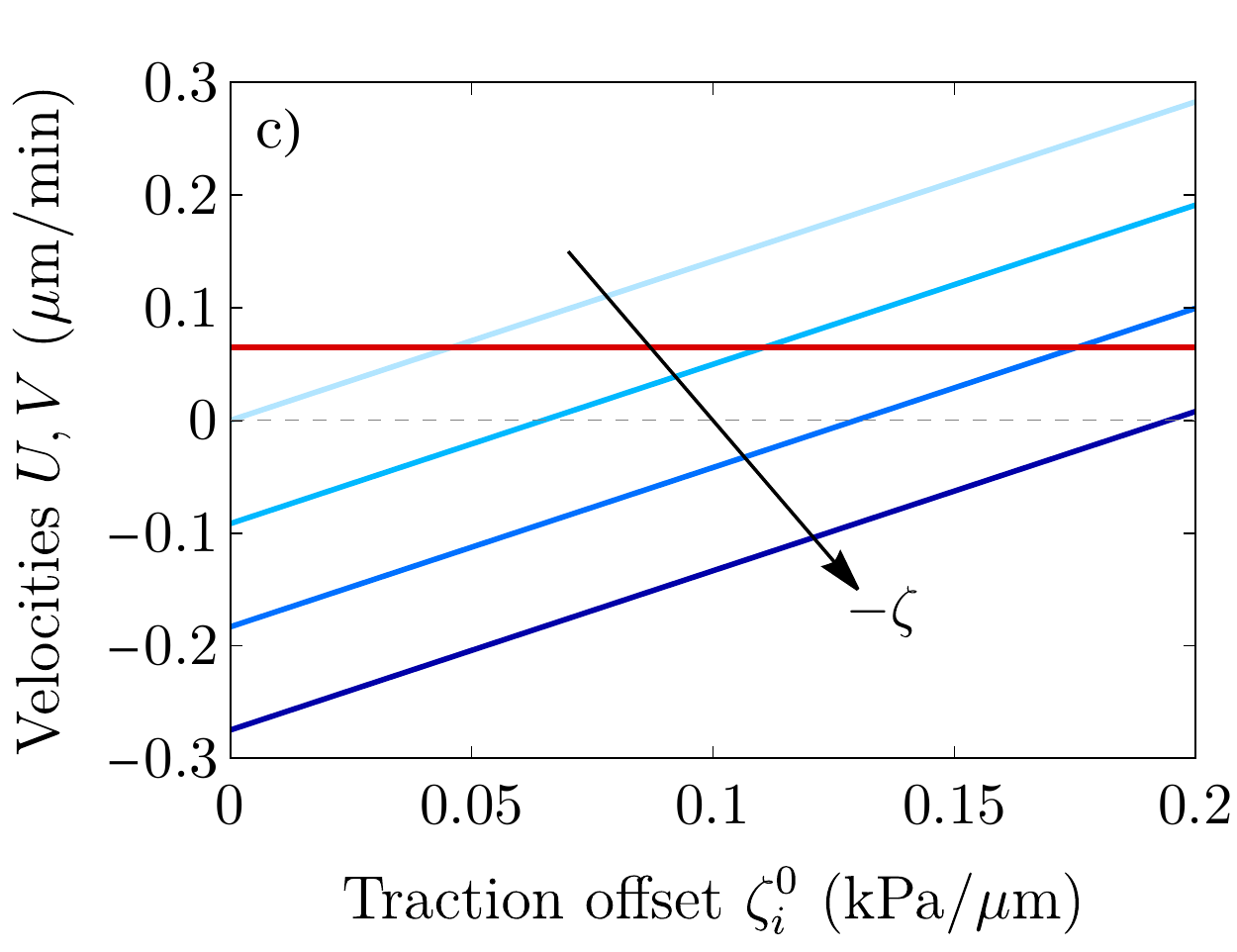}\hfill
  \includegraphics[width=.485\columnwidth, trim={0.8cm 0 0 0},clip]{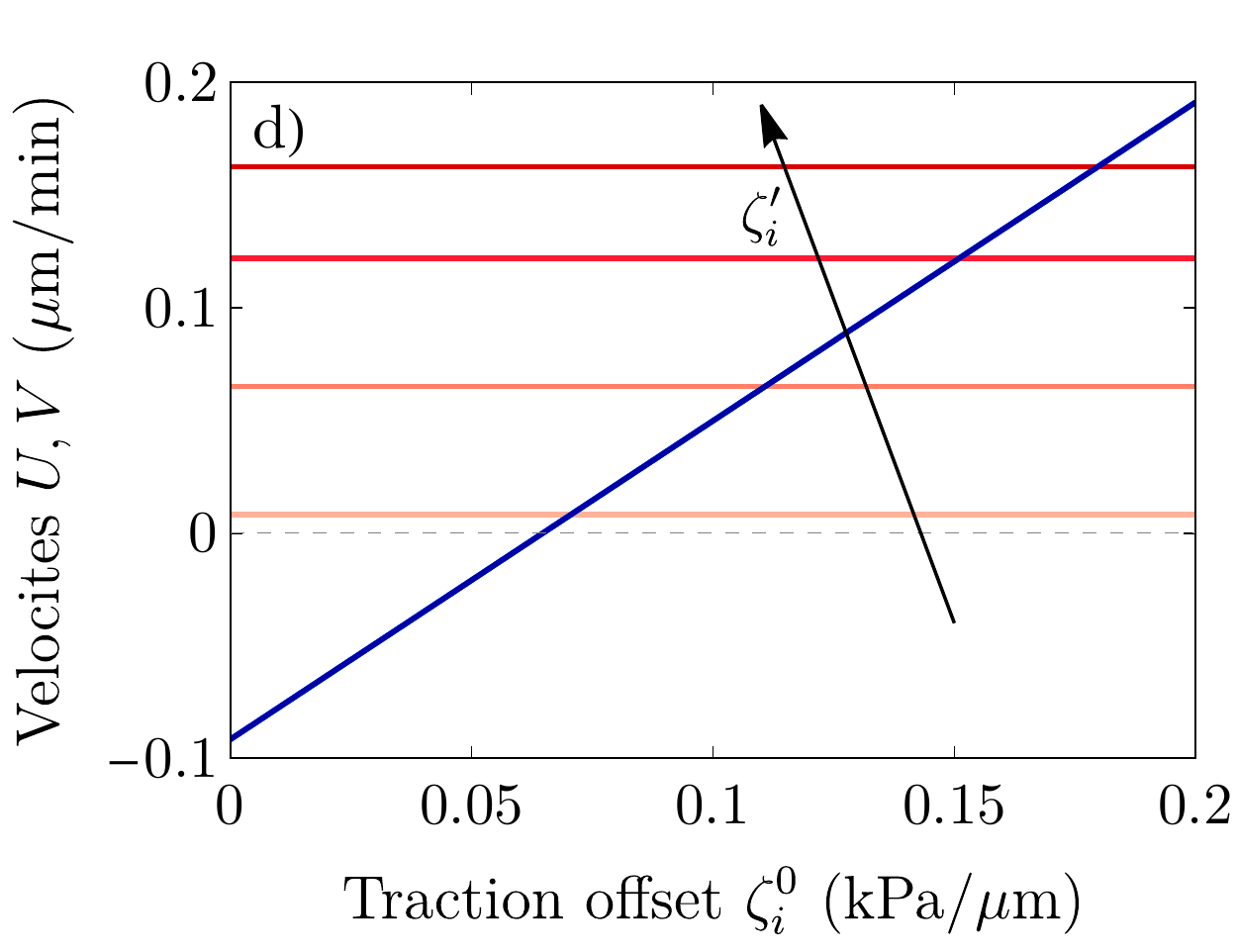}
  \caption{Plots of the full expressions for $V$ (blue curves) and $U$ (red curves) as a function of the traction offset $\zeta_i^0$ for the parameter values in \cref{tab param} but changing (a) the tissue size $L = 40, 100, 200, 350$ $\mu$m, (b) the hydrodynamic length $\lambda = 100, 200, 300, 450$ $\mu$m, (c) the contractility $-\zeta = 0, 20, 40, 60 $ kPa, and (d) the traction gradient $\zeta_i' = 10^{-5}, 8\times 10^{-5}, 1.5\times 10^{-4}, 2\times 10^{-4}$ kPa/$\mu$m$^2$.}
  \label{fig vvsTrac_param}
\end{figure}

The repertoire of dynamical behaviours contained in the model as a function of parameters is quite rich. Spreading and center-of-mass velocities can be plotted against monolayer size (\cref{fig vvsL0_param}) and traction offset (\cref{fig vvsTrac_param}), which are two quantities that can be easily varied and controlled in experiments \cite{Sunyer2016}. Importantly, in addition to being independent of traction offset, the durotactic velocity $U$ does not depend on the contractility either, which is a parameter that is more difficult to infer from experiments, and is assumed to be uniform throughout the system. An increase in either monolayer size $L$ or traction gradient $\zeta_i'$ implies an increase of the difference of local tractions at the edges, $\zeta_i^+-\zeta_i^-$, and thus an increase in durotactic velocity $U$. The spreading velocity $V$, which is independent of the traction gradient $\zeta_i'$, increases with the monolayer size $L$, the screening length $\lambda$, and the traction offset $\zeta_i^0$, and it decreases with the contractility $|\zeta|$.

\begin{figure}[tb!]  
  \includegraphics[width=.50\columnwidth, trim={0 0cm 0 0},clip]{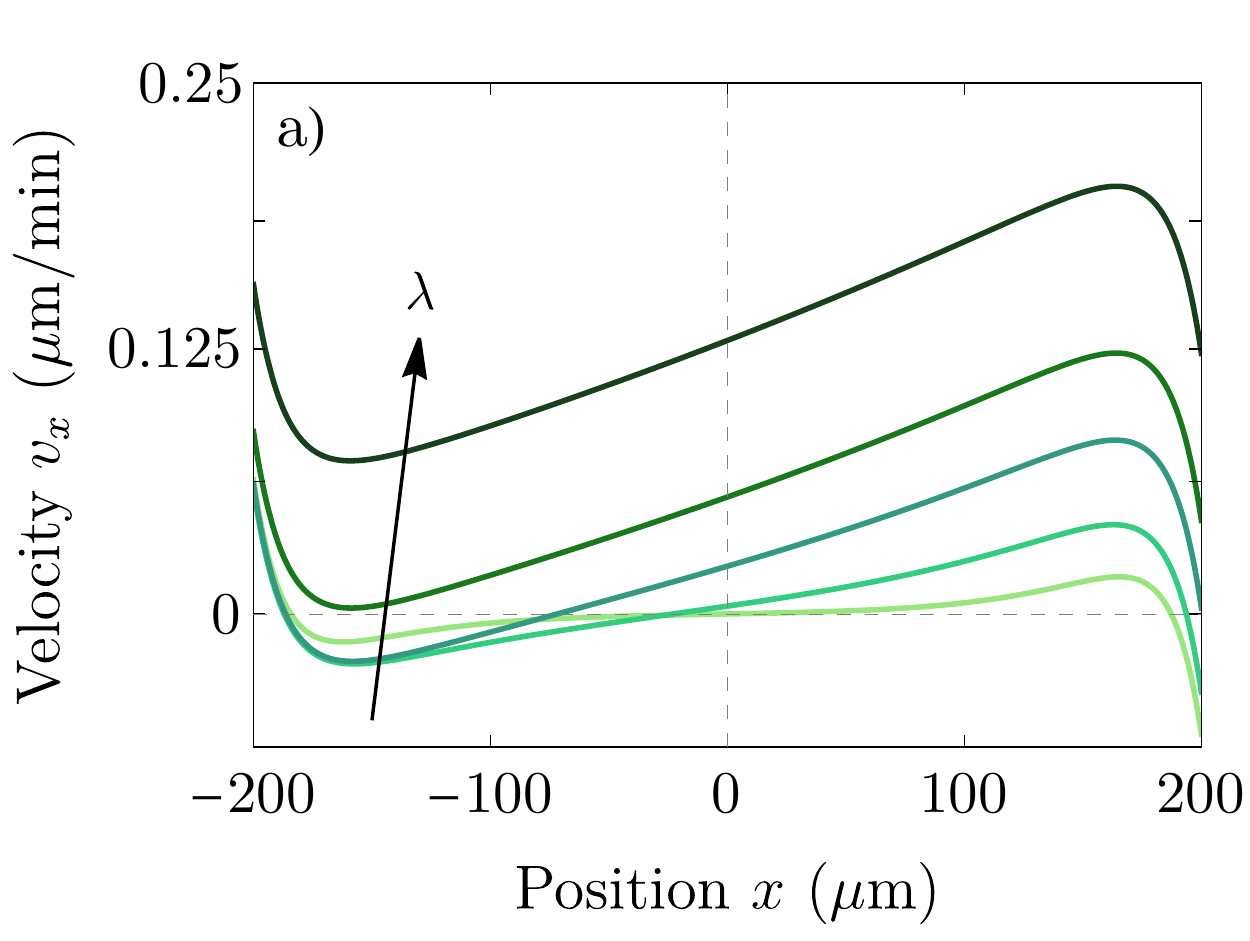}\hfill
  \includegraphics[width=.48\columnwidth, trim={0 0 0 0},clip]{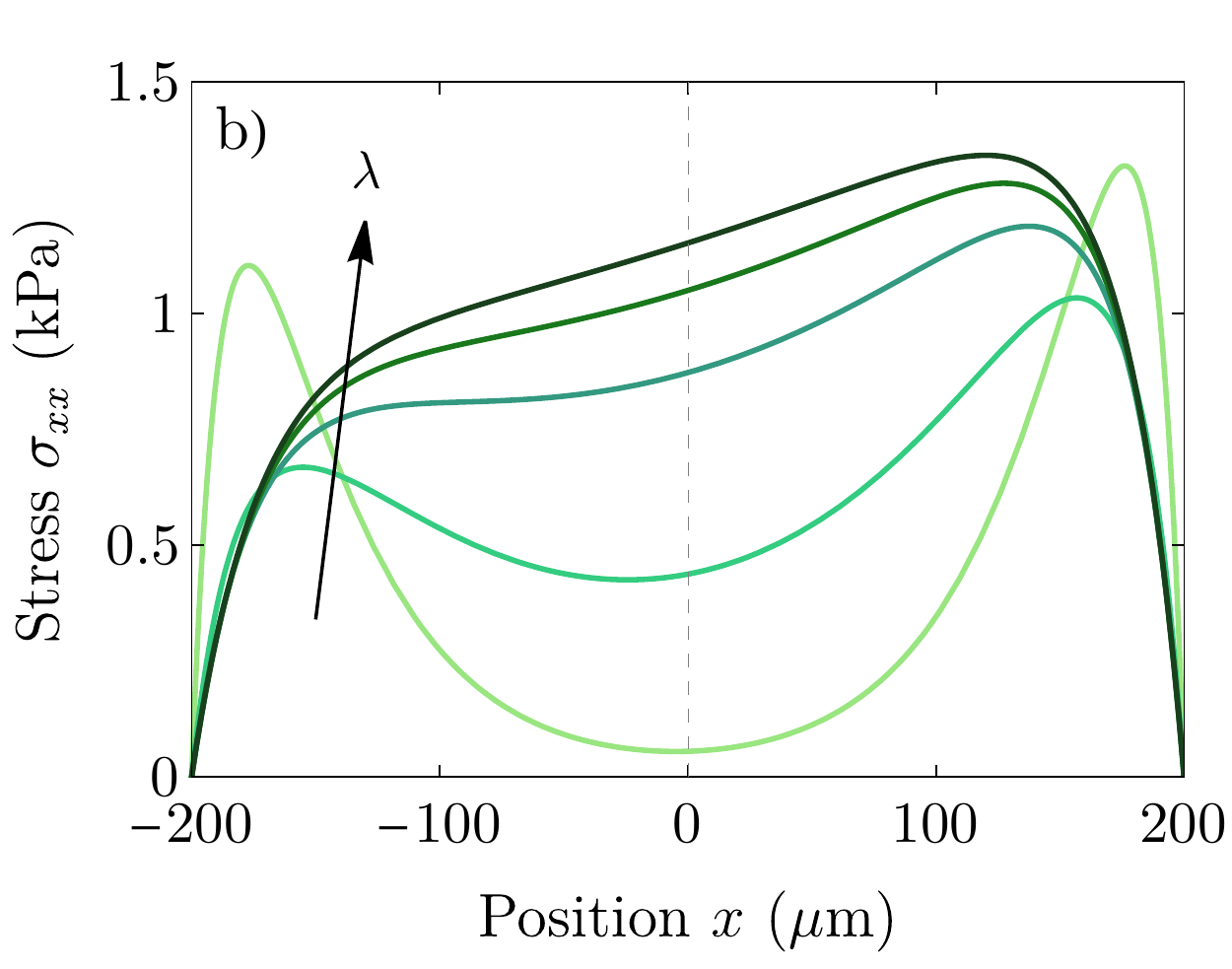}\hfill
  \caption{Velocity (a) and stress (b) profiles for a linear traction profile, for the parameter values in \cref{tab param} and changing $\lambda = 40,100,200,300,450$ $\mu$m. Note that with the values of $v_x$ at the tissue edges we would obtain the spreading and durotactic velocities in \cref{fig vvsTrac_param}b, for $\zeta_i^0=0.05$ kPa/$\mu$m.} 
  \label{fig velstress_profile_grad}
\end{figure}

The velocity and stress profiles, plotted in \cref{fig velstress_profile_grad} for a range of $\lambda$, are qualitatively similar to those of the uniform-stiffness substrate (\cref{fig velstress_profile_const}), except that they become asymmetric due to the stiffness gradient.

\subsection{Linear friction and saturation profiles}
\label{sec:3_results_durotaxis_friction}

\begin{figure}[tb!] 
  \centering
  \includegraphics[width=0.49\columnwidth]{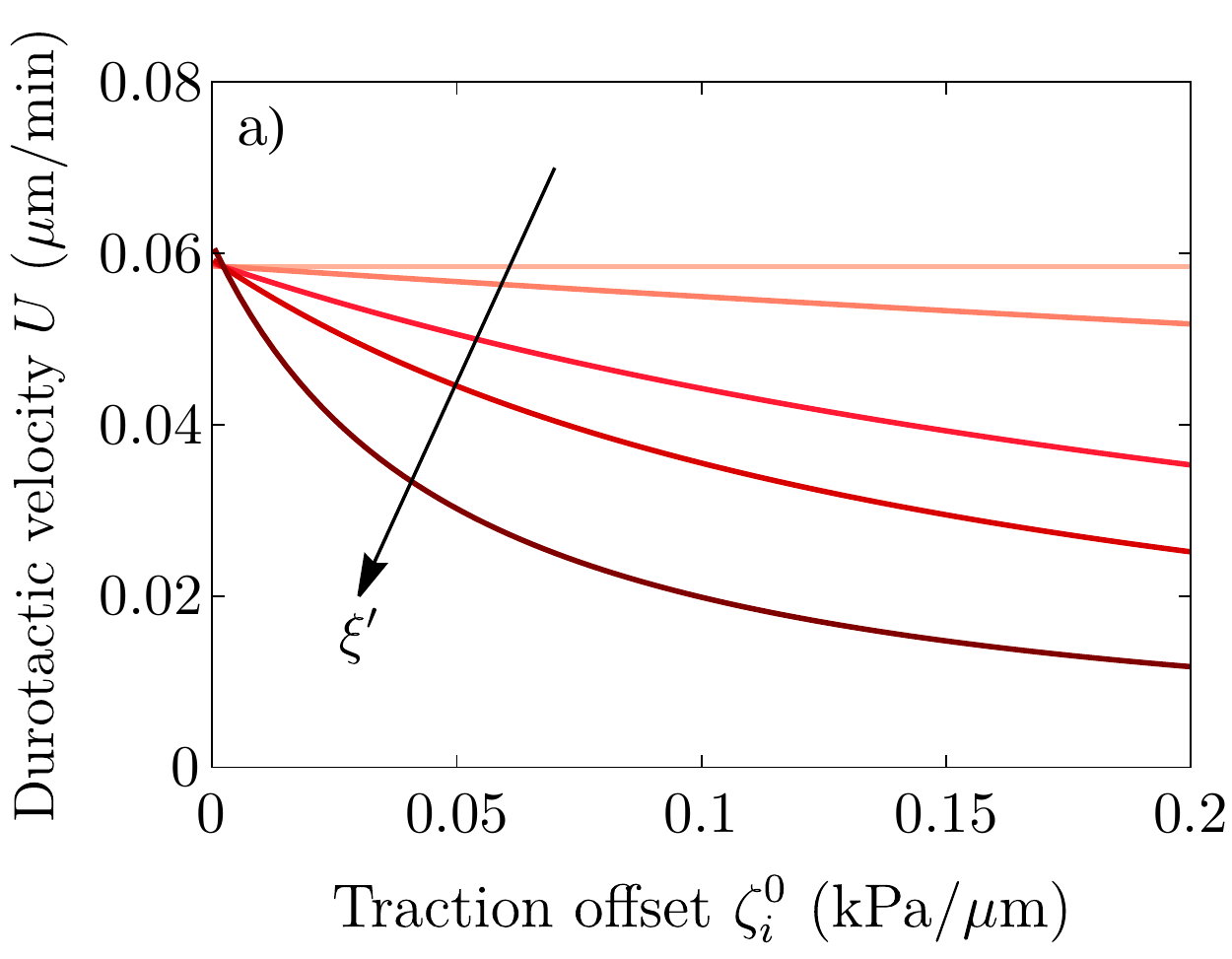} \hfill
  \includegraphics[width=0.49\columnwidth]{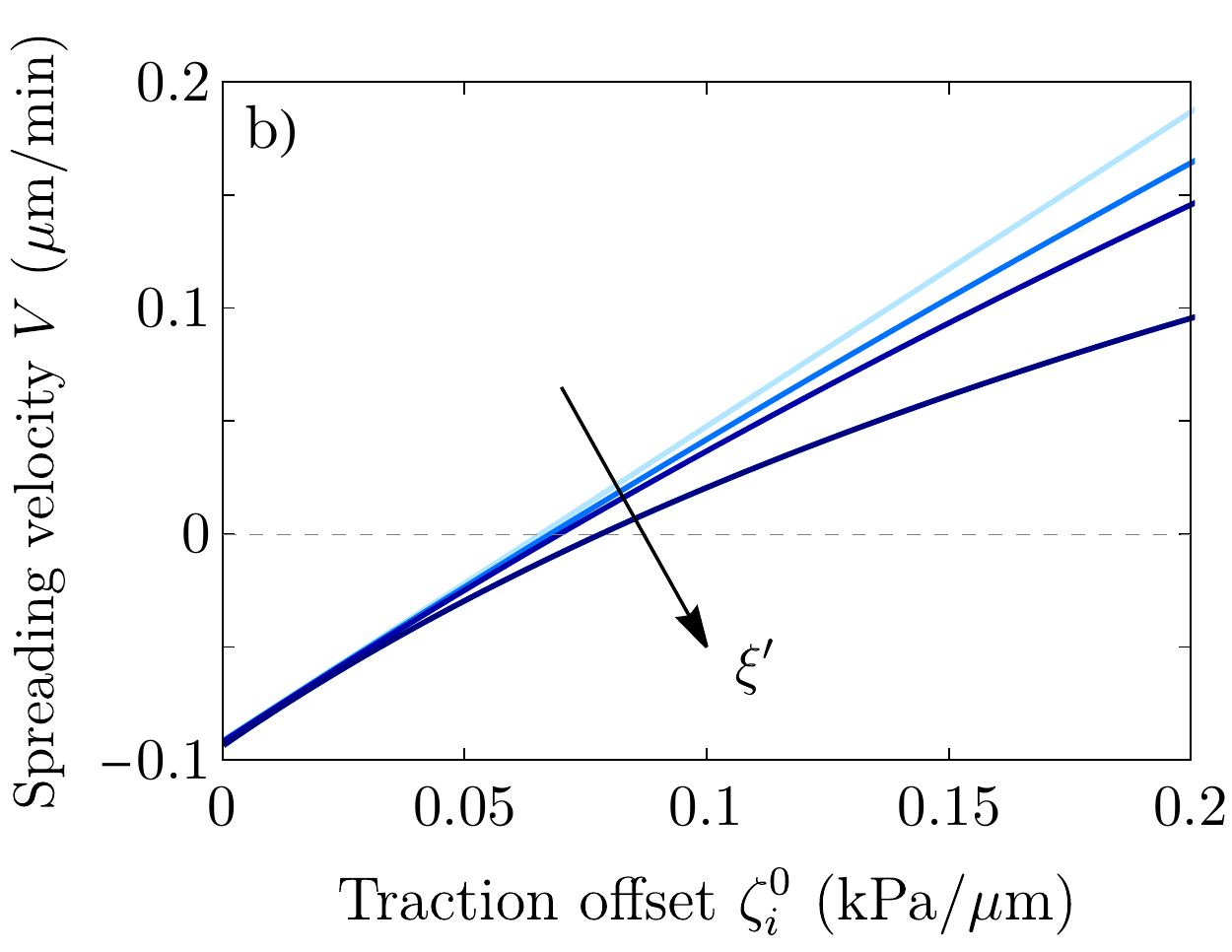}
  \caption{Durotactic velocity $U$ (a) and spreading velocity $V$ (b) when there is a positive gradient of the friction coefficient, for parameters in \cref{tab param}, varying the friction gradient $\xi' = 0, 10^{-4}, 5\times 10^{-4}, 10^{-3}, 3\times 10^{-3}$ kPa s/$\mu$m$^3$, and taking a stiffness offset $\xi_0 = 2 $ kPa s/$\mu$m$^2$.}
  \label{fig vvsTrac_fricgrad}
\end{figure}

In this section we relax the restriction of a uniform friction coefficient. This is a more realistic situation since both active traction and passive friction rely on the dynamics of cell-substrate adhesion molecules \cite{Oriola2017}, and hence they both depend on substrate stiffness. Previous works indeed support that cell-substrate friction increases with substrate stiffness \cite{Walcott2010,Saez2010,Trichet2012,Gupta2015}. To illustrate the role of this effect on tissue durotaxis, and for the sake of simplicity, we consider a linear friction increase $\xi(x) = \xi_0 + \xi' x$, with $\xi' \neq 0$. The problem with space-dependent $\xi$ can no longer be solved analytically. Solving \cref{eq main} numerically with a finite-difference method, we find that the center-of-mass velocity now decreases with the traction offset (\cref{fig vvsTrac_fricgrad}). This is because now larger traction correlates with larger friction, which leads to smaller velocities. Accordingly, the spreading velocity grows more slowly with traction offset than in the uniform-friction case.

As already mentioned, the use of linear profiles is particularly convenient from a theoretical point of view since it avoids introducing too many parameters. However, to obtain a more realistic description and to compare with experimental data, other profiles may be more appropriate. A simple feature that can be implemented in the model is the fact that the increase of traction and friction with substrate stiffness must eventually saturate. Then, following Ref. \cite{Alert2019a} and the references therein, we consider profiles of the form 
\begin{equation} \label{eq saturation-profiles}
    \zeta_i(x) = \zeta_i^{\infty}\frac{E(x)}{E(x)+E^*},  \quad \xi(x) = \xi^{\infty}\frac{E(x)}{E(x)+E^*},
\end{equation}
in terms of the spatially-varying Young modulus $E(x)$ of the substrate. For \emph{in vitro} experiments such as those of Ref. \cite{Sunyer2016}, a simple choice is to prepare the substrate with a linear stiffness profile $E=E_0+E'(x-X)$. Numerical results for this more general case are qualitatively similar to those in \cref{fig vvsTrac_fricgrad}. However, at high stiffness, the saturation of traction and friction makes the tissue dynamics approach those of the uniform-stiffness case, with vanishing durotactic velocity $U$. The approach to this durotaxis-free regime at high stiffness is controlled by the new parameters $\zeta_i^\infty$, $\xi^\infty$, $E^*$ and $E'$ introduced in \cref{eq saturation-profiles}.

\section{Time evolution in a traction gradient}
\label{sec:4_evolution}

\subsection{General case}
\label{sec:4_evolution_general}
For a given set of parameters, $\eta$, $\zeta$, $L_c$, initial conditions $X_0$, $L_0$, boundary conditions, and imposed profiles $\zeta_i(x)$ and $\xi(x)$, our model supplies a velocity profile $v(x)$ as the solution of the equation 
\begin{equation}
\left( 2\eta \partial_x^2 - \xi(x) \right)v(x) = \left( 2 \zeta p(x) \partial_x   - \zeta_i(x) \right) p(x),  \label{eq main2}
\end{equation}
where $p(x)=p(x;X,L,L_c)$ is given by \cref{eq polarity_sol}. The position of the center of mass $X(t)$ and the cluster size $L(t)$ satisfy the differential equations
\begin{align}
\dot{X} & =\frac{v(X+L) + v(X-L)}{2}, \\
\dot{L} & =\frac{v(X+L) - v(X-L)}{2}.
\end{align}
As $X$ and $L$ evolve, however, the cell cluster is visiting different regions of the substrate, so the profiles $\zeta_i(x)$ and $\xi(x)$ used to solve \cref{eq main2} are changing with time. For instance, in the case of a linear traction profile, $\zeta_i(x) = \zeta_i^0 + \zeta_i' (x-X) $, the traction offset changes with time according to $\zeta_i^0(t)=\zeta_i^0(0) + \zeta_i'(X(t)-X_0)$. In the rest of this section we focus on this case with and take no friction gradient ($\xi'=0$).

\subsection{Uniform traction gradient}
\label{sec:4_evolution_traction_grad}
Since the durotactic motion is toward increasing traction ($U>0$ for $\zeta_i'>0$), the local traction offset $\zeta_i^0$ increases with time. In a uniform traction gradient, the durotactic velocity $U$ is insensitive to the local traction offset. Therefore, the increasing traction offset does not lead to an increasing durotactic speed. However, $U$ depends also on the monolayer size $L$, which may grow or decay according to the sign of the spreading velocity $V$. In general, $U$ increases monotonically with $L$. Therefore, if $L$ is large enough, the monolayer spreads and then the center-of-mass velocity $U$ increases during the evolution as $L$ increases. As a conclusion, monolayer spreading produces increasingly faster durotaxis.

For small enough $L$, the monolayer initially contracts ($V<0$). However, as the tissue moves toward stiffer regions, it may reach values of traction that are large enough to change the sign of $V$ and produce a transition to spreading. In this case, the evolution of $L$ is non-monotonic in time, corresponding to initial contraction followed by spreading. Finally, if $L$ is even smaller, the durotactic velocity $U$ may not be sufficient to reach sufficiently large values of traction to reverse the sign of $V$ to produce spreading. In this case, the monolayer contracts (dewets) completely into a three-dimensional spheroid. 

The asymptotic behaviour of the system at long times is thus either indefinite expansion or the collapse. In both situations, the model is no longer adequate as additional physics will take over at long times. In the case of asymptotic spreading, even if the traction forces do not saturate, other forces such as elastic forces may eventually slow down and even suppress the spreading as discussed below. In the case of monolayer retraction and collapse with $L\rightarrow 0$, the quasi-twodimensional description breaks down and a more elaborate treatment of the three-dimensional structure of the tissue becomes necessary. 

\begin{figure}[tb!]
  \includegraphics[width=.5\columnwidth, trim={0 0cm 0 0},clip]{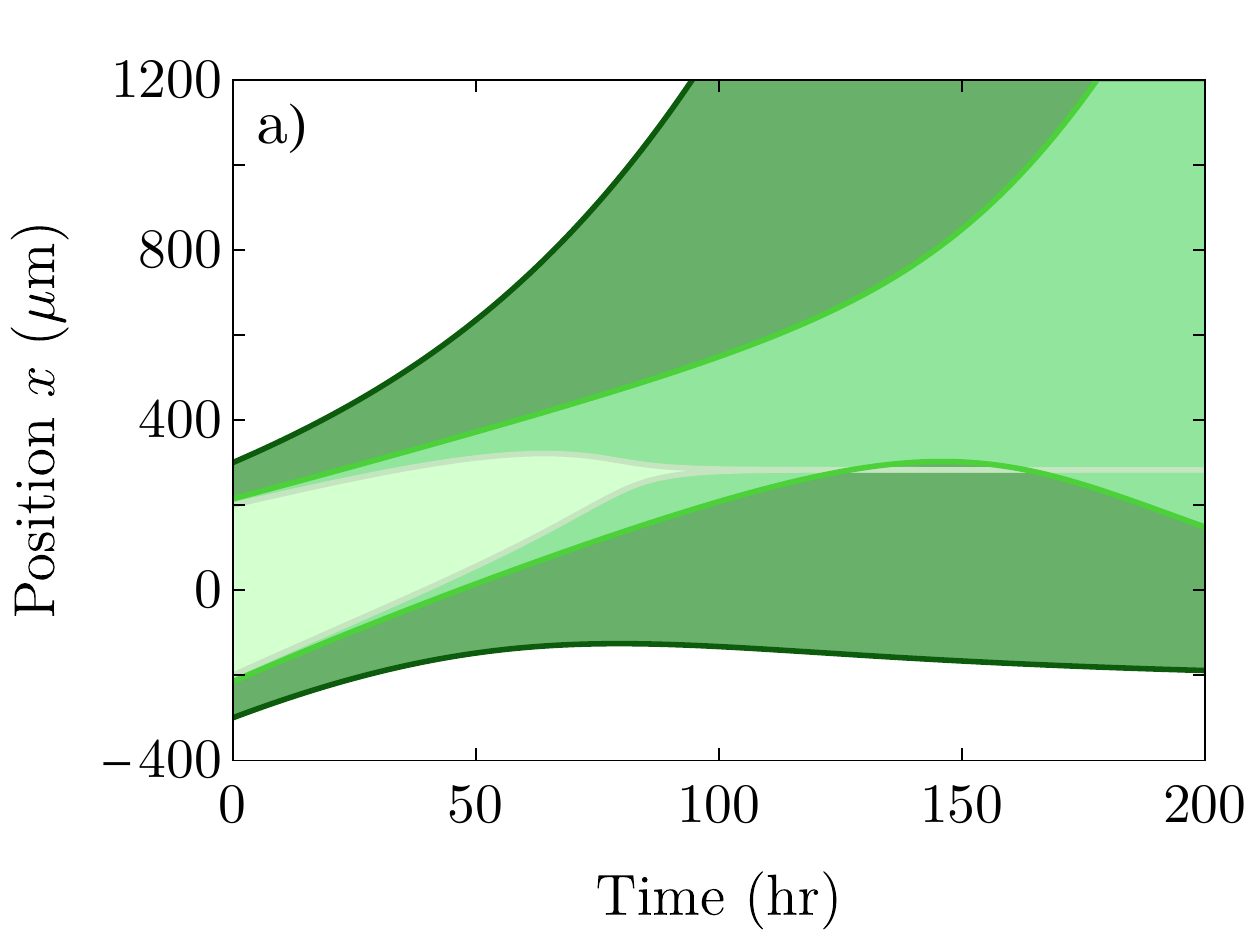}\hfill
  \includegraphics[width=.46\columnwidth, trim={0cm 0cm 0 0},clip]{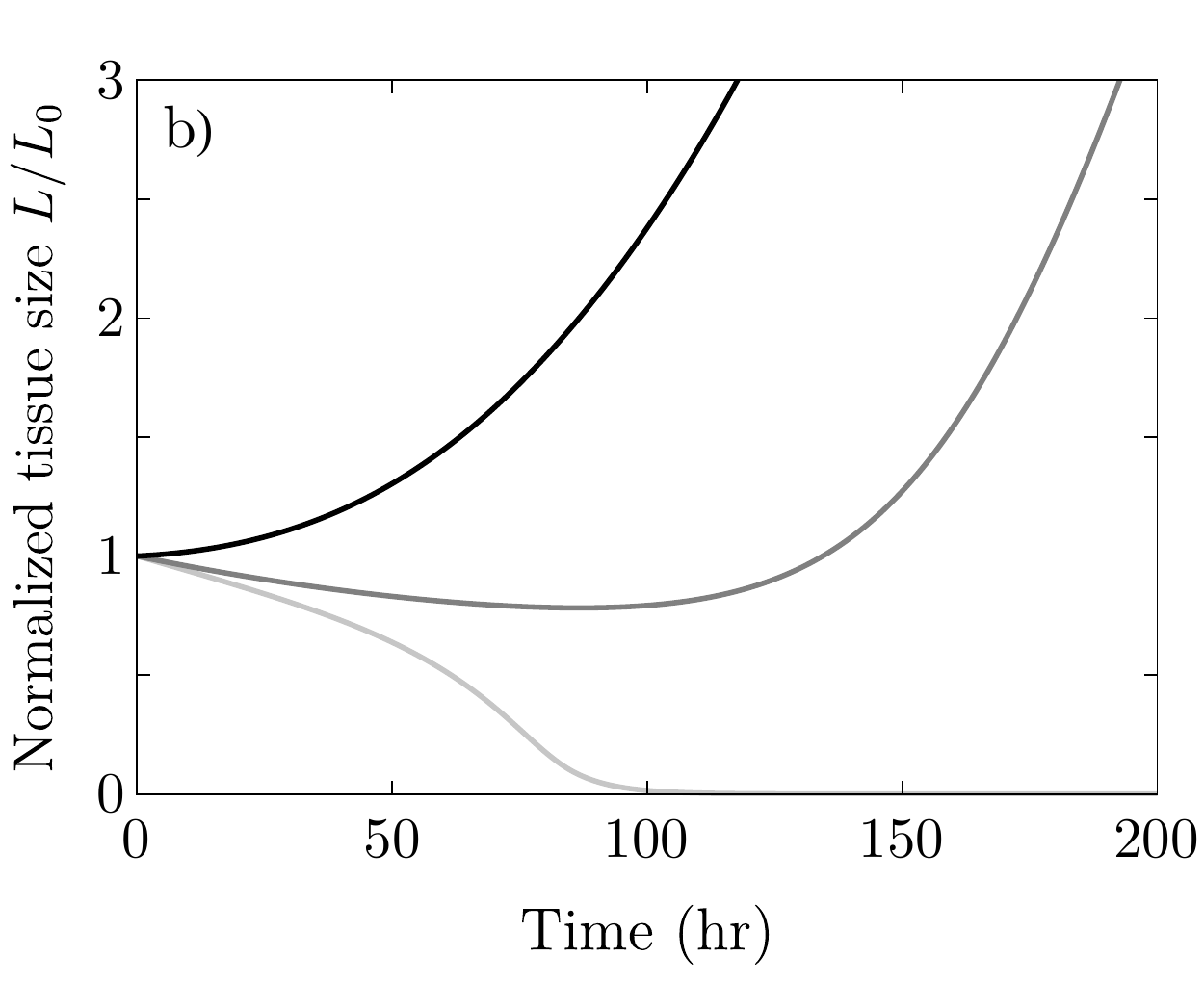}
  \includegraphics[width=.5\columnwidth]{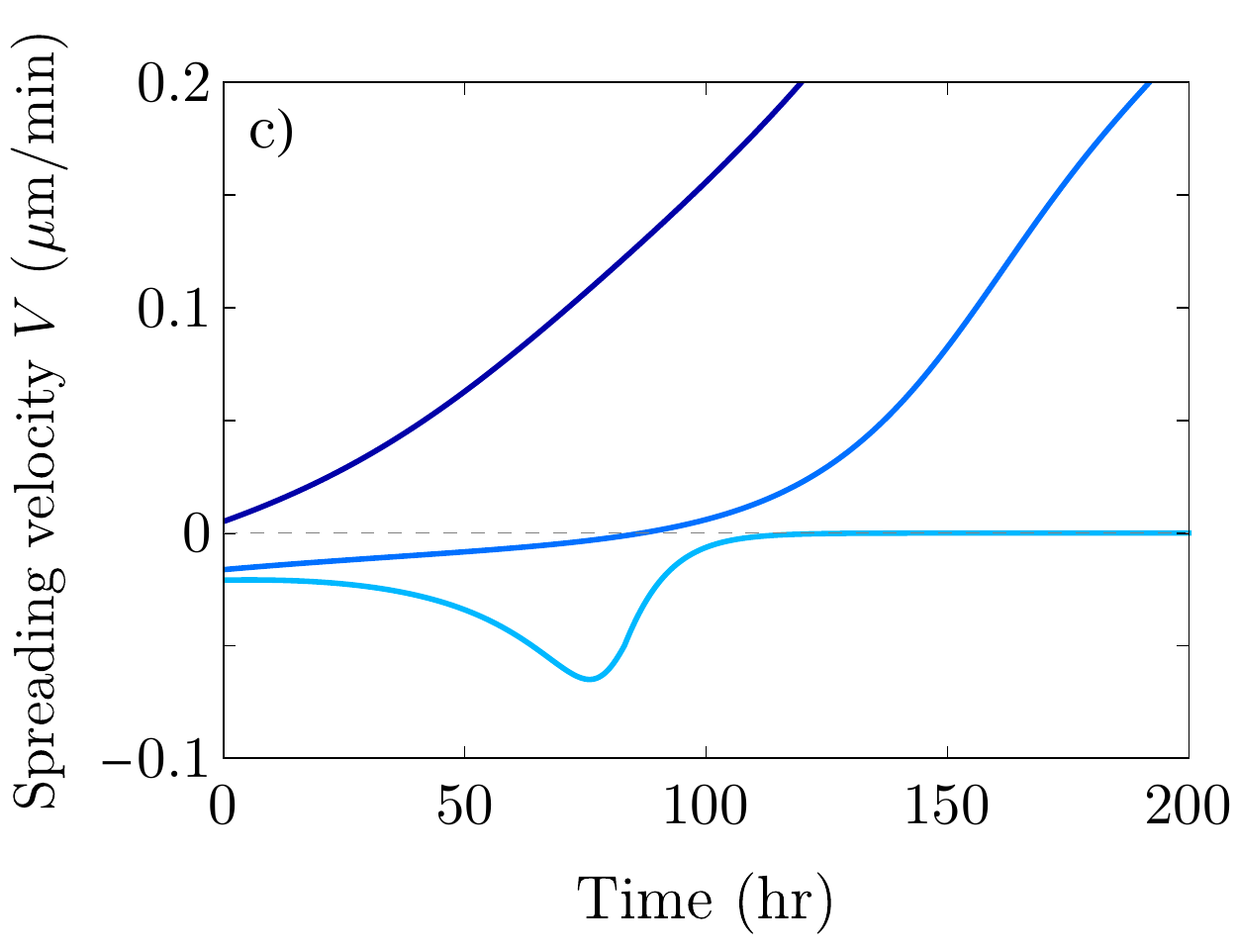}\hfill
  \includegraphics[width=.5\columnwidth, trim={0cm 0 0 0},clip]{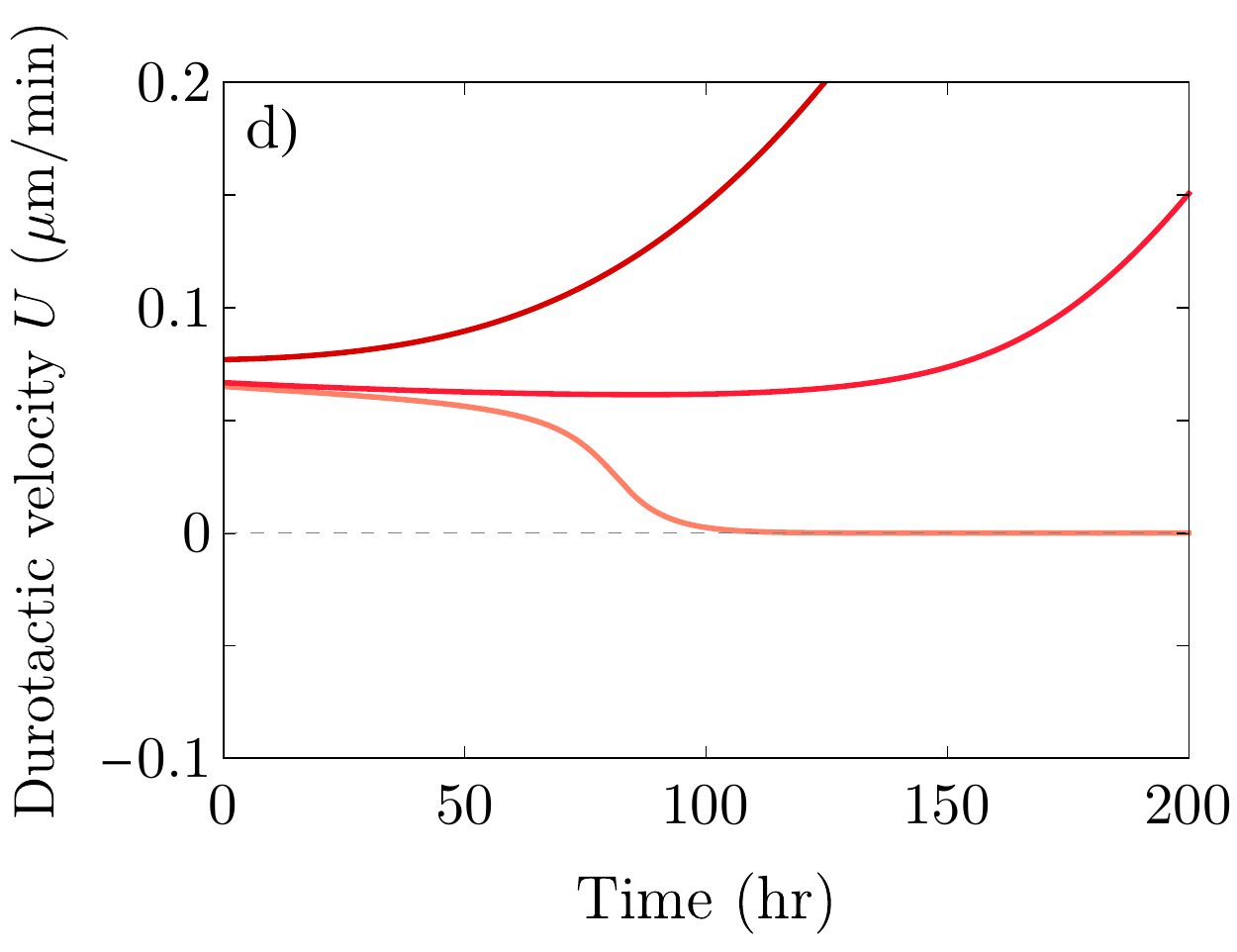}
  \caption{Time evolution of a cell monolayer on a traction gradient. (a) Position of the monolayer edges $x_\pm(t)$, filling the area between them to represent the tissue width. (b) Monolayer width divided by its initial value. (c) Spreading velocity. (d) Center-of-mass velocity (d). In each plot, curves from lighter to darker show three different examples with $L_0 = 200,215$ and $300$ $\mu$m, which are characteristic of the three different dynamical regimes. The initial center-of-mass position is $X_0=0$ $\mu$m in all three cases, the traction gradient $\zeta_i'$ is uniform, the friction is uniform ($\xi'=0$), and other parameter values are those in \cref{tab param}. Here, $L^* = 276.35$ $\mu$m and $L^{*c} \approx 213$ $\mu$m. The tissue contracts when the normalized $L$ and $U$ decrease and $V<0$, whereas the tissue expands when $L$ and $U$ increase and $V>0$. The regime with initial contraction and later expansion presents an almost constant durotactic velocity $U$ and tissue width $L$. The corresponding edge velocities together with $U$ and $V$ in each case are shown in \cref{fig evolution_edges}.}
  \label{fig evolution}
\end{figure}

For a given set of parameters, the tissue width $L$ controls the transitions between the three possible spreading dynamics. For $L \geq L^*$, the monolayer expands for all times $V(t)>0$. Here, $L^*$ is the critical size for spreading on uniform substrates ($V(L^*)=0$), which we discussed before. An explicit and exact expression for $L^*$ is given by equating \cref{eq const_full,eq grad_full_vs} to zero. For intermediate values of $L$, $L^{*c}< L < L^*$, the monolayer initially contracts ($V<0$) and later expands ($V>0$) indefinitely. Finally, for $L \leq L^{*c}$, the monolayer contracts for all times ($V(t)<0$). The three regimes are illustrated in \cref{fig evolution}.

\subsection{Some generalizations of the model}
\label{sec:4_generalizations}
\begin{figure}[tb!]
  \centering
  \includegraphics[width=0.8\columnwidth]{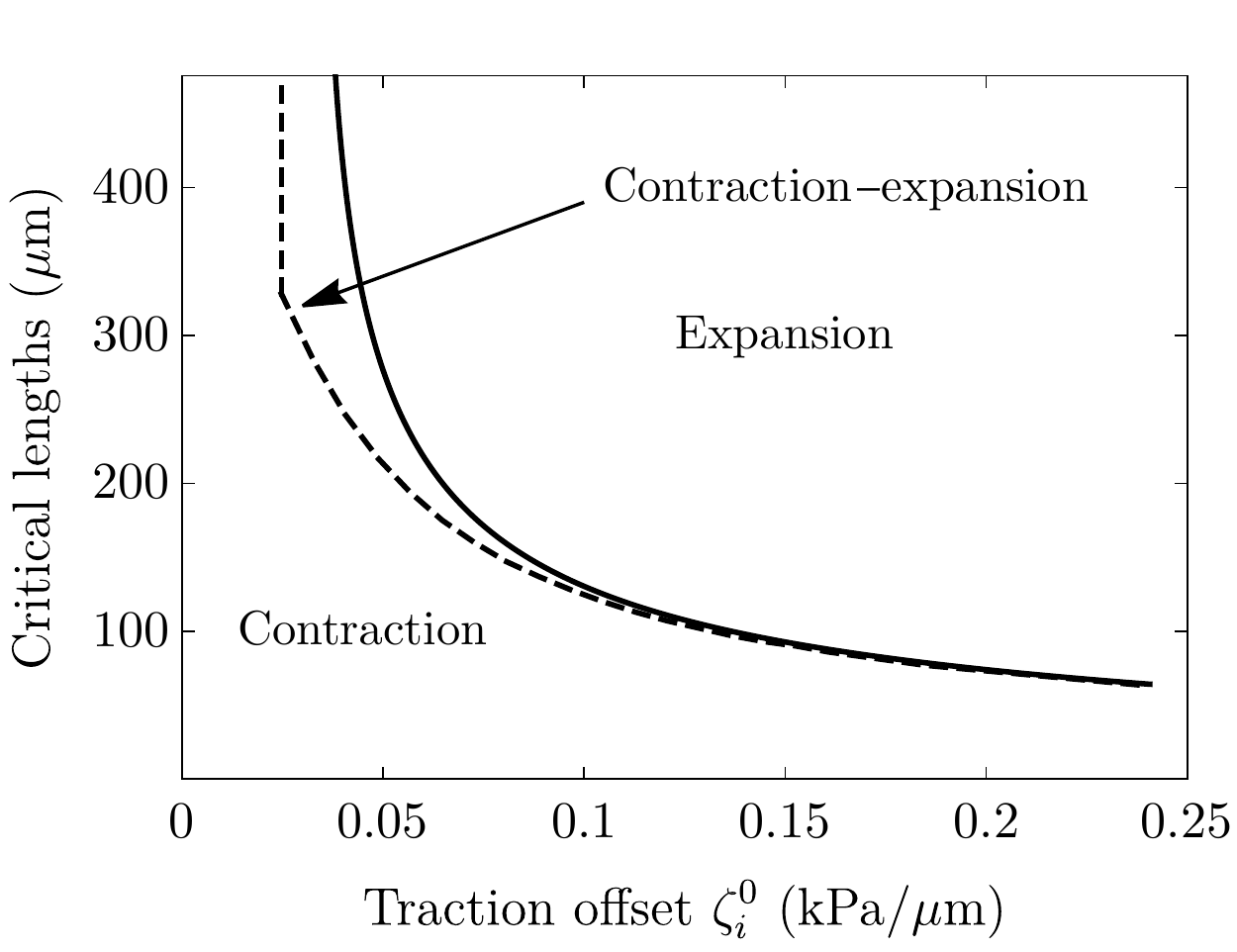}
  \caption{Critical lengths defining the three spreading regimes as a function of traction offset. The solid curve corresponds to $L^*$, which is the solution of $V(L^*) =0$ obtained from \cref{eq grad_full_vs}. The dashed curve corresponds to $L^{*c}$, which defines the length below which the monolayer contracts for all times ($V(t)<0$). The region between both curves defines the intermediate contraction-expansion regime. Parameter values are given in \cref{tab param}.}
  \label{fig L0crit}
\end{figure}

More general profiles of traction and friction can also be used for studying the time evolution. As long as the profiles are both monotonically increasing, the qualitative behaviour is similar. The three spreading regimes discussed above, separated by the critical lengths $L^*$ and $L^{*c}$, still exist, but their expressions and values change. For the case of uniform traction gradient $\zeta_i'$ and no friction gradient ($\xi'=0$), given the initial monolayer size $L_0$, we predict the critical lengths as a function of the traction offset, as shown in \cref{fig L0crit}. 

As mentioned above, the two possible asymptotic behaviours of the monolayer dynamics are not particularly interesting. This is because the traction profile can not grow indefinitely, and other physical effects will either stop the extreme stretching of the cells in the case of spreading or enable the formation of a three-dimensional cell aggregate in the case of contraction. The latter will not be considered here because it requires essential modifications of the model that are deferred to future work. However, different effects may be easily introduced in our current model either to slow down the indefinite spreading of the cluster size or even to stop it.

The first possibility is to introduce of an effective surface tension $\gamma$ at the tissue edge, as already mentioned in \cref{sec:2_model_eq}. The introduction of this surface tension can be understood if we interpret our 1d model as an approximation for a circular monolayer of radius $L$. This surface tension slows down the spreading process, less effectively for larger monolayers. For contracting monolayers, surface tension accelerates the contraction. 

A less trivial but more determinant modification is to introduce an effective elastic force that prevents excessive cell stretching. This type of force has been introduced at a phenomenological level for single cells to favor a characteristic cell size. It was used for instance in Refs. \cite{Recho2013a,Hennig2020} in effective 1d models for single-cell motility. Such an elastic force, together with the Young-Laplace pressure drop due to surface tension $\gamma$, can be implemented in the following boundary condition for the stress:
\begin{equation} 
\sigma_\pm = -\frac{\gamma}{L} - k\frac{(L - L^r)}{L^r}.
\end{equation}
Here, $k$ is an elastic constant, and $L^r$ is a characteristic size of the cell monolayer, which is proportional to the number of cells if the cell size is somehow regulated.

\begin{figure}[tb!]  
  \includegraphics[width=.52\columnwidth, trim={0 0cm 0 0},clip]{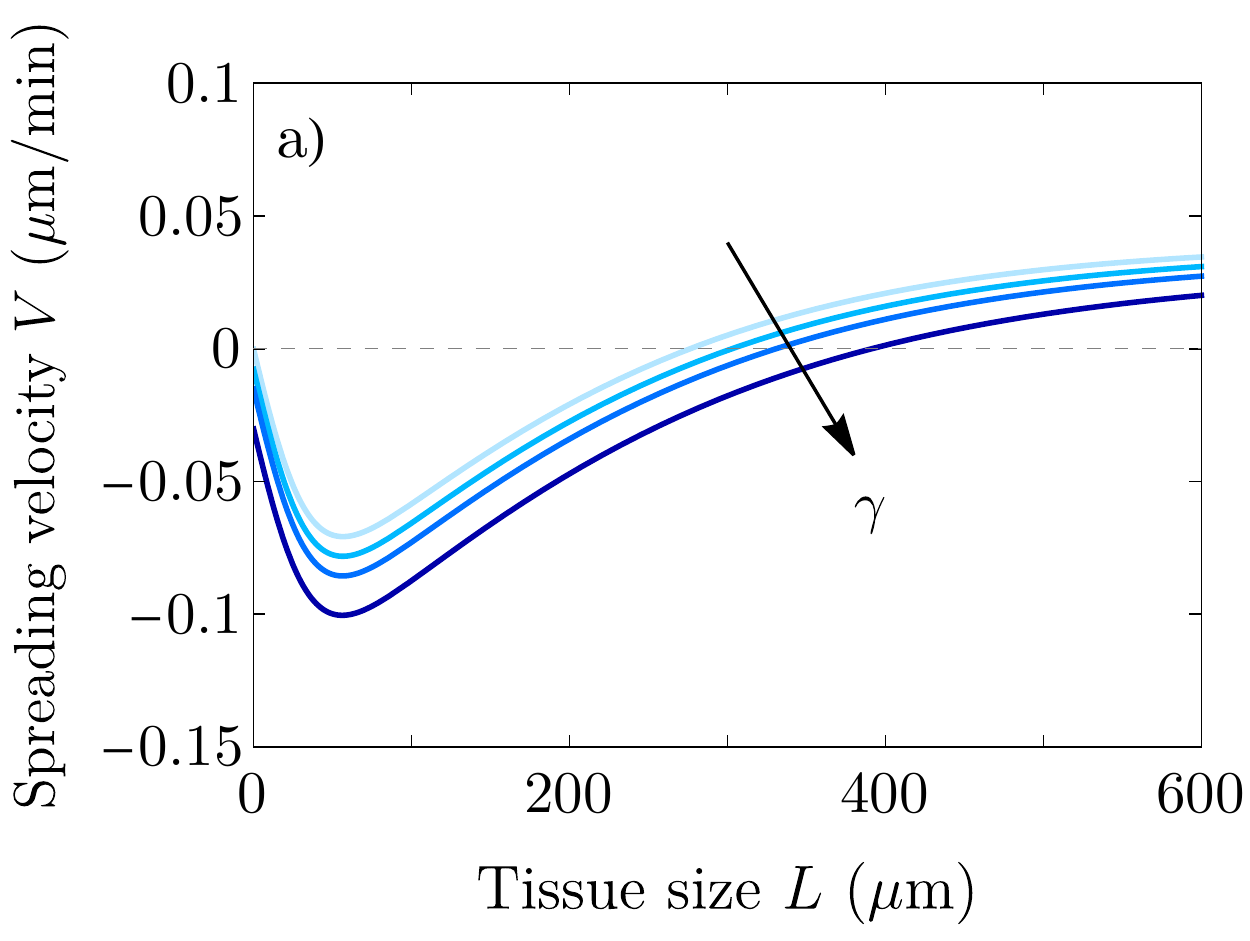}\hfill
  \includegraphics[width=.46\columnwidth, trim={1.35cm 0 0 0},clip]{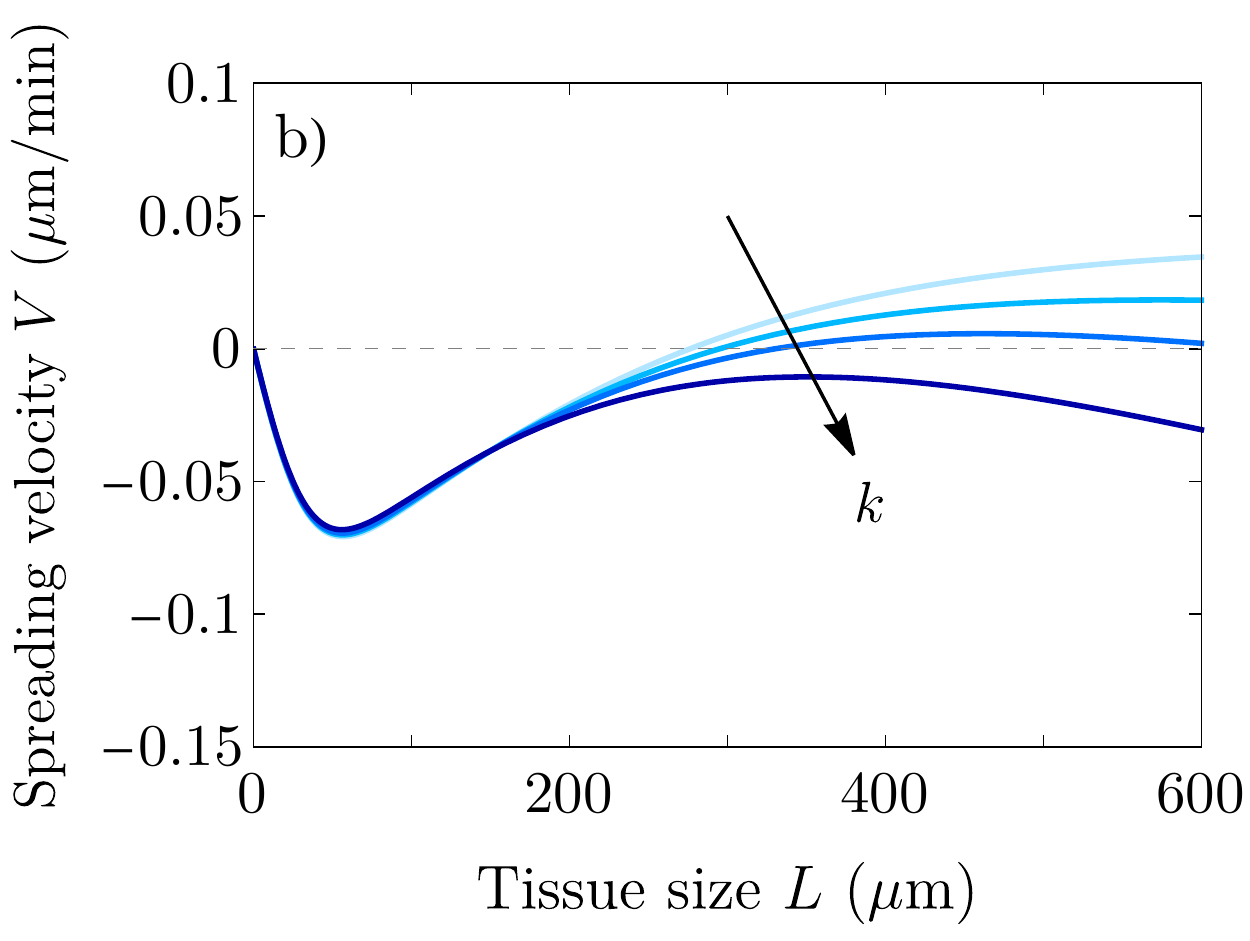}\hfill
  \caption{Spreading velocity $V$ as a function of tissue size $L$ for the parameter values in \cref{tab param} but with (a) surface tension $\gamma = 0,20,40,80$ kPa·$\mu$m ($k=0$), and (b) elastic constant $k = 0,0.05,0.1,0.2$ kPa ($L^r=150$ $\mu$m and $\gamma=0$). Here, to showcase its effects, we take values of $\gamma$ larger than what is measured experimentally for cell aggregates (\cref{tab param}). Respectively, we take $k$ comparable to $\zeta_i L_c \approx \sigma$.} 
  \label{fig vvsL0_gammak}
\end{figure}

\begin{figure}[tb!] 
  \centering
  \includegraphics[width=.52\columnwidth, trim={0 0cm 0 0},clip]{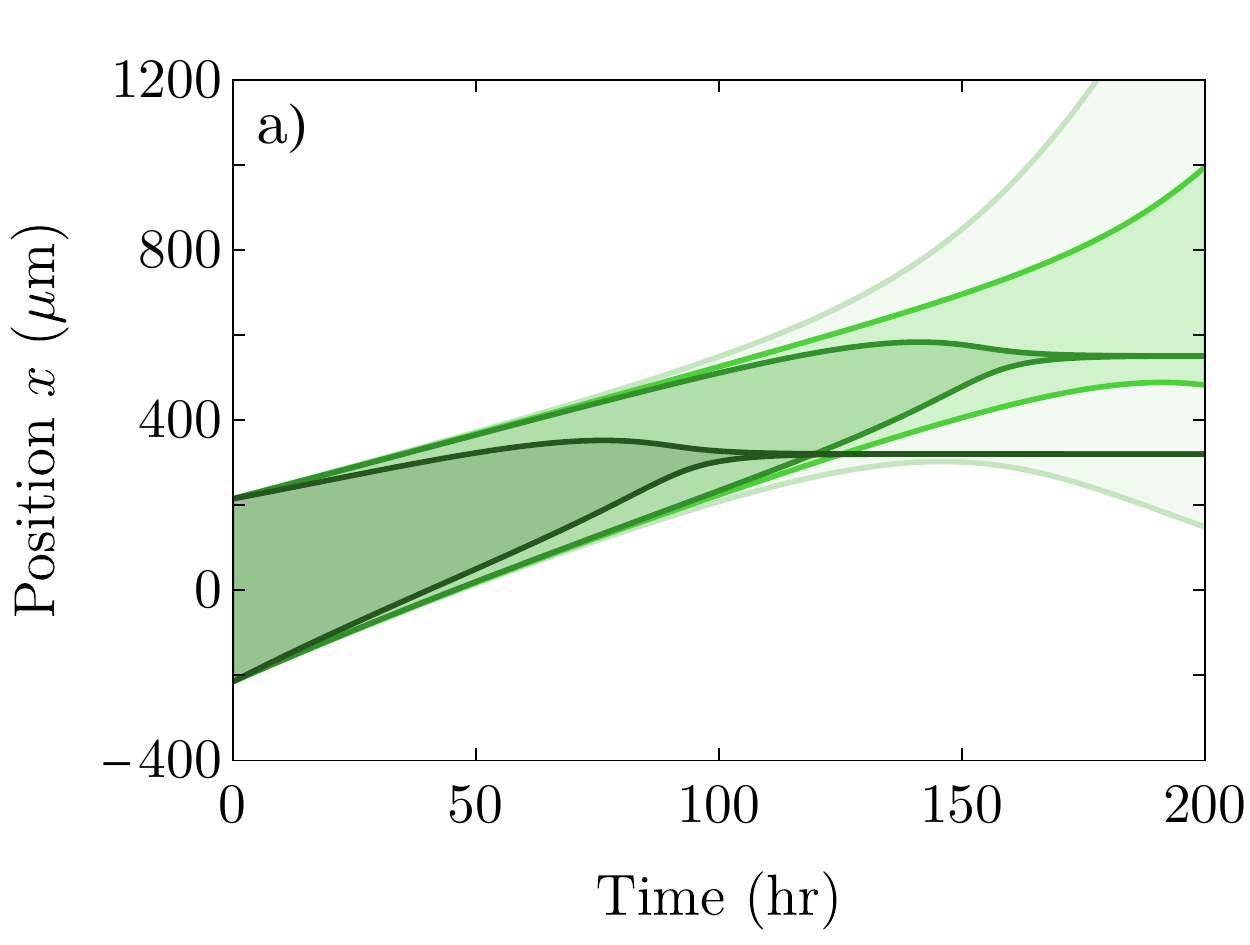}\hfill
  \includegraphics[width=.48\columnwidth, trim={1cm 0 0 0},clip]{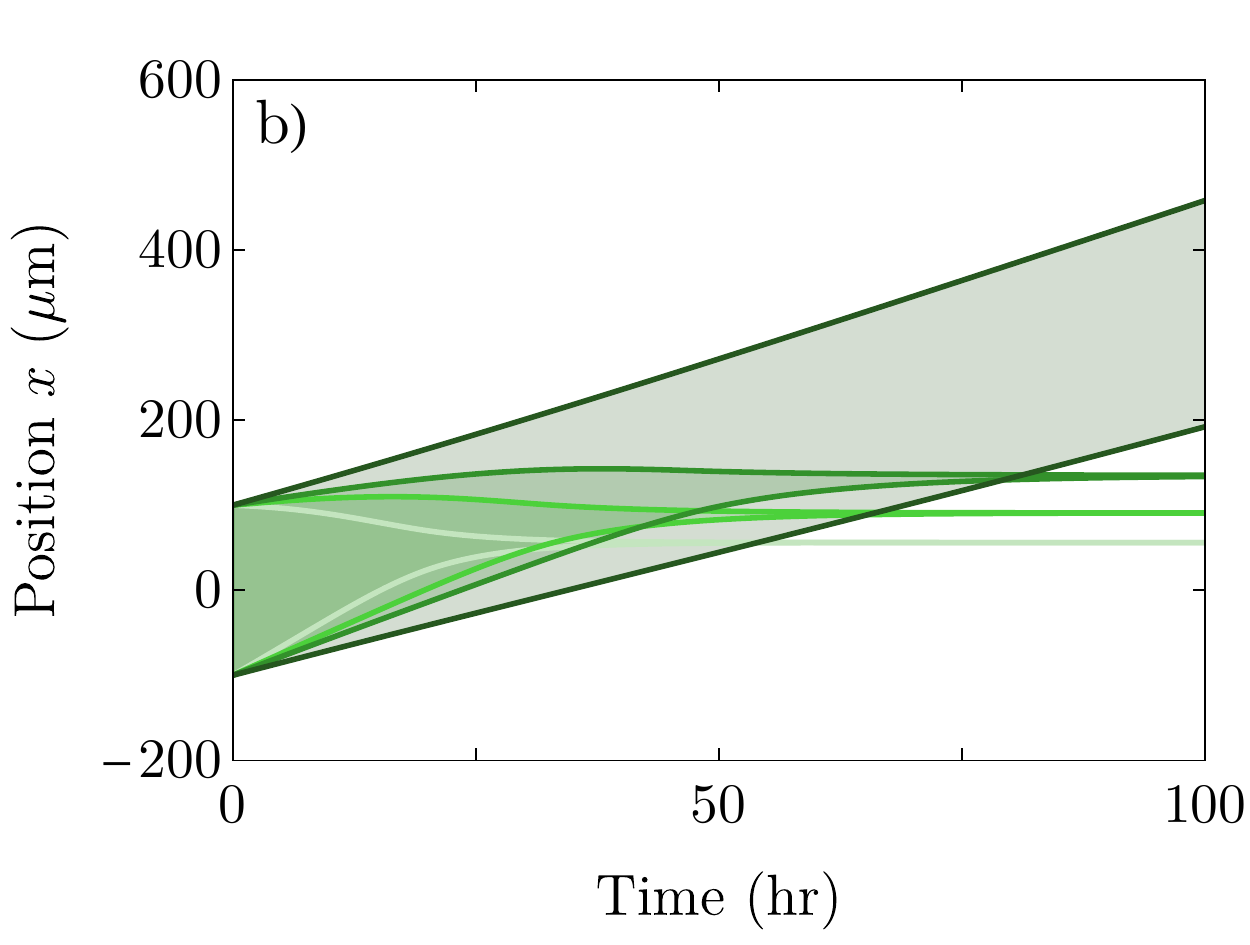}
  \includegraphics[width=.52\columnwidth, trim={0cm 0 0 0},clip]{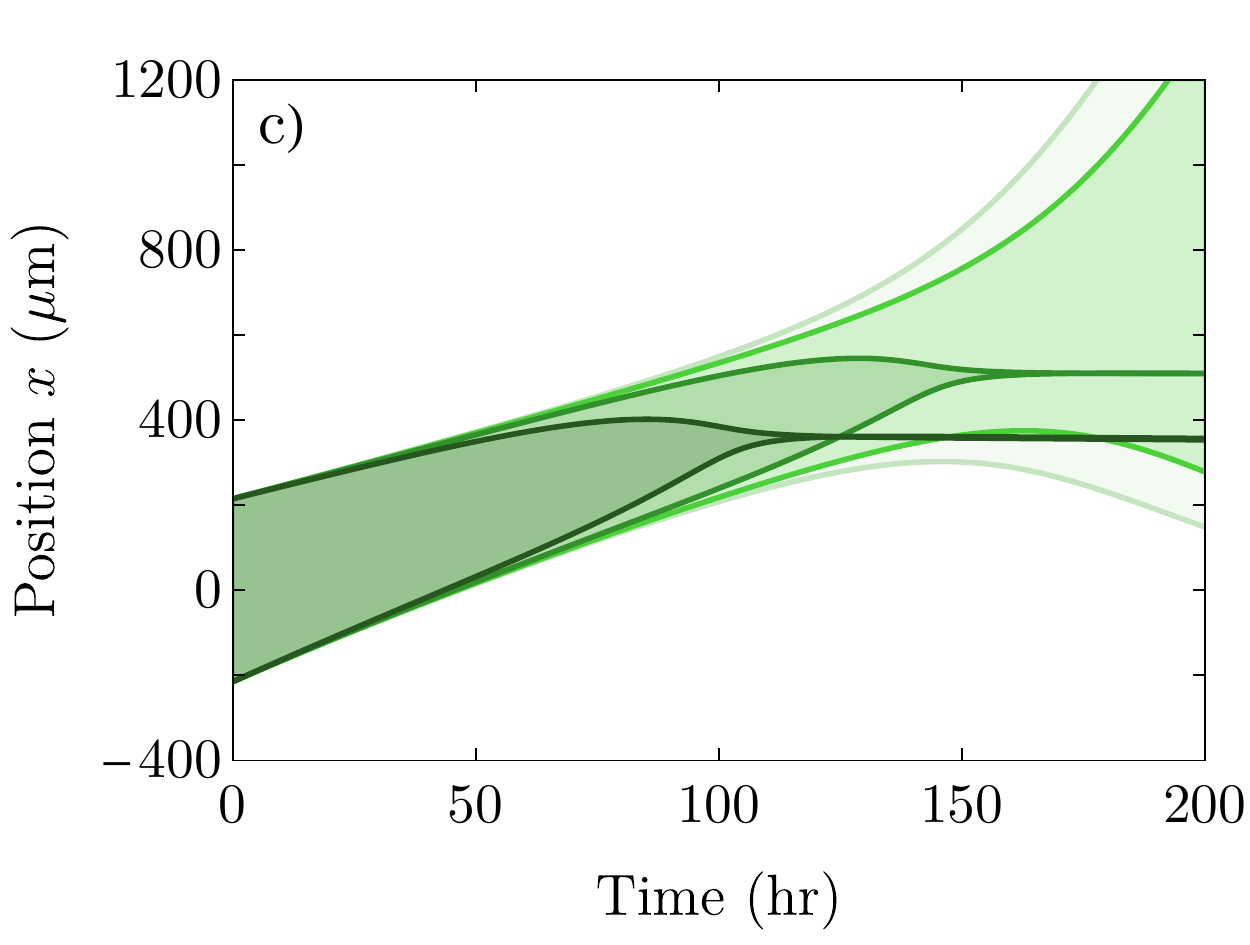}\hfill
  \caption{Examples of monolayer spreading dynamics to illustrate the effect of elasticity $k$ (first row) and surface tension $\gamma$ (second row). In each plot, curves of the same color show the evolution of the position of the edges $x_\pm(t)$, filling the area between them to represent the tissue width. In the first row, $\gamma=0$ and $L^r=150$ $\mu$m. In (a), the initial size is $L_0 = 215$ $\mu$m and $k =  0,0.03,0.05,0.5$ kPa. In (b), $L_0 = 100$ $\mu$m and $k = 0,2,3,5$ kPa. The elastic constant $k$ increases from lighter to darker curves. In (c), $k=0$ and only the $L_0 = 215$ $\mu$m case is shown with $\gamma = 0,1,3,10$ mN/m, which also increases from lighter to darker green. Other parameter values are those in \cref{tab param}.} 
  \label{fig evolution_gammak}
\end{figure}

For a uniform traction gradient $\zeta_i'$ and no friction gradient ($\xi'=0$), the center-of-mass velocity $U$ turns out to be independent of both surface tension and elasticity (see \cref{app grad_substrate}). The spreading velocity $V$, however, is affected, respectively giving
\begin{align} 
    V(\gamma) &= V(\gamma=0) -\frac{\gamma}{L} \frac{\lambda}{2\eta}\tanh{\left(\frac{L}{\lambda}\right)}, \label{eq vs_gamma} \\
    V(k) &= V(k=0) -k\frac{L-L^r}{L^r} \frac{\lambda}{2\eta}\tanh{\left(\frac{L}{\lambda}\right)} \label{eq vs_k}
\end{align}
when either surface tension or elasticity are added. The surface tension $\gamma$ always decreases the spreading velocity, while the elastic term contributes with a different sign depending on whether the monolayer size is larger or smaller than $L^r$, always in the direction of approaching the reference value $L^r$ (\cref{fig vvsL0_gammak}). Both effects affect the spreading dynamics, changing for instance the critical lengths, but the phenomenology and qualitative evolution of the monolayer typically remains unchanged. However, for large $k$ and $L>L^r$  (\cref{fig evolution_gammak}a), a monolayer that starts spreading can change to contraction. In this case, similar to surface tension (\cref{fig evolution_gammak}c), elasticity slows down expansion and accelerates contraction. On the other hand, if $L<L^r$ (\cref{fig evolution_gammak}b), elasticity accelerates expansion and slows down contraction, although only very large values of $k$ ($k \gg T L_c \sim \sigma$), presumably not biologically possible, enable a transition from contraction to expansion.

\section{Conclusions}   
\label{sec:5_conclusions} 

In this work we have presented a comprehensive study of collective cell durotaxis based on a continuum model of cell monolayers as a two-dimensional active fluid on a gradient of substrate stiffness. The stiffness gradient affects tissue dynamics through a spatially-dependent traction and friction coefficients $\zeta_i(x)$ and $\xi(x)$. We analytically solve the model in a one-dimensional setup. The tissue dynamics is characterized by two main observables: the velocity of the center of mass and the spreading velocity. For the simple case of a uniform traction gradient $\zeta_i'$ and uniform friction ($\xi'=0$), the spreading velocity is exactly the same as that for the uniform substrate case, so the spreading behavior is independent of the existence of a traction gradient. The velocity of the center of mass, instead, is finite and proportional to $\zeta_i'$. Therefore, the cell monolayer performs durotaxis as long as the traction is a monotonically increasing function of the substrate stiffness. These conclusions are locally valid for more general traction profiles provided that the gradient does not change significantly over the monolayer width.

We have analyzed the durotactic dynamics as a function of physical parameters, for example discussing the wet and dry limits that result from comparing the monolayer size $L$ to the hydrodynamic screening length $\lambda$. For broad ranges of values and profiles physical parameters, and for different boundary conditions, we have characterized the different regimes that result from combining states of spreading and contraction with states of interface wetting and dewetting. All of them give rise to durotactic motion. The durotactic velocity increases with both the traction gradient $\zeta_i'$ and the monolayer size $L$, as seen in Ref. \cite{Escribano2018a}. Moreover, for uniform $\zeta_i'$ and uniform friction, the durotactic velocity is independent of the contractility $\zeta$ and the traction offset $\zeta_i^0$. Therefore, the same monolayer placed at different positions along the stiffness gradient would have the same durotactic velocity but different spreading dynamics.

However, for non-uniform friction ($\xi'\neq 0$), as well as for traction and friction profiles that saturate with stiffness, the model predicts lower velocities for larger stiffness offsets, thus recovering the results in Ref. \cite{Sunyer2016,Escribano2018a,Gonzalez-Valverde2018}. At high stiffness, parameter saturation makes the system asymptotically approach the dynamics on uniform substrates, with vanishing durotaxis. The spreading velocity increases with the traction offset $\zeta_i^0$, the monolayer size $L$, and the hydrodynamic length $\lambda$, but it decreases with contractility.

In addition to the predictions for local durotaxis and spreading, we have discussed the temporal evolution of a monolayer along the stiffness gradient as it changes its position and size. We have identified three regimes for the evolution of monolayer size. Large monolayers spread indefinitely, small monolayers contract indefinitely, and monolayers in an intermediate size range display a non-monotonic evolution whereby they switch from contraction to spreading at a finite time. These three regimes are separated by two critical lengths, which we determined analytically in some simple cases, and we illustrated numerically for more general situations. We also discussed the effect of additional physical ingredients such as surface tension and elastic forces that oppose large deformations of the tissue. We have shown that they typically slow down the expansion and accelerate the contraction.

Our model is relatively simple and strongly predictive, so it could be tested in experiments and used to infer parameter values from experimental data. It could also guide the design of further experiments on collective durotaxis. Nevertheless, the model has obvious limitations: It is restricted to cell monolayers, and it is unrealistic in its long-time behaviour. Addressing these limitations would require to include additional physics such as effects from the three-dimensional, multiple-layer structure resulting from monolayer contraction, additional forces to prevent indefinite spreading, and other ingredients such as cell proliferation. All these generalizations of the model are deferred to future work. Finally, it is a question of great interest to elucidate to what extent a purely mechanical description, with no need to invoke biochemical signaling, can account for the observed phenomenology in different forms of collective cell migration.

\begin{acknowledgements}
I.P-J. acknowledges support from an FPU19/05492 grant from the Spanish Government.
R.A. acknowledges support from the Human Frontier Science Program (LT000475/2018-C).
J.C. acknowledges support by MIN-ECO (project PID2019-108842GB-C21) and Generalitat de Catalunya (project 2017-SGR-1061).
\end{acknowledgements}

\section*{Author contributions}
R.A. and J.C. proposed the model. All authors contributed to the development and interpretation of the theory. I.P-J performed analytical calculations and numerical integration. I.P-J and J.C. wrote the paper.

\vfill

\bibliography{All}
\onecolumngrid

\clearpage

\appendix
\addcontentsline{toc}{section}{Appendices}
\section*{Appendices}

\section{Uniform substrate} \label{app const_substrate}
The solution to \cref{eq main} for a constant traction $\zeta_i$ and friction $\xi$ is obtained assuming a normal component of the stress in the boundaries due to two different effects: an effective surface tension $\gamma$ (interpreting our 1d model as an approximation for a circular cluster of radius $L$), and an effective elastic stiffness $k$, accounting for a mean-field-type linear elastic interaction as in Ref. \cite{Recho2013a} that prevents the tissue from excessive stretching, being $L^r$ the reference length (justification more extended in \cref{sec:4_generalizations}). Thus, 
\begin{equation}
    \sigma_{\pm} = -\frac{\gamma}{L}-k\frac{L-L^r}{L^r} \longrightarrow \partial_x v |_{x_{\pm}}= \frac{1}{2\eta} \left(\zeta - \frac{\gamma}{L}-k\frac{L-L^r}{L^r} \right).
\end{equation}
Without loss of generalisation we can take $X=0$ ($U=0$). The solution for the velocity profile reads
\begin{align} \label{eq const_full_profile}
    v(x) =& \frac{\lambda}{2\eta}\Bigg[ \left(\zeta -\frac{\gamma}{L} - k\frac{L-L^r}{L^r} + \frac{\lambda^2 L_c \zeta_i}{\lambda^2-L_c^2}\coth{(L/L_c)} - \frac{2\zeta\lambda^2}{4\lambda^2 - L_c^2}\left( 2 + \csch^2{(L/L_c)}\right) \right)\frac{\sinh{(x/\lambda)}}{\cosh{(L/\lambda)}}
    \nonumber \\ & + \frac{\lambda L_c}{\sinh{(L/L_c)}}\left(\frac{\zeta}{4\lambda^2-L_c^2}\frac{\sinh{(2x/L_c)}}{\sinh{(L/L_c)}}-\frac{\zeta_i L_c}{\lambda^2-L_c^2}\sinh{(x/L_c)} \right)\Bigg].  
\end{align}
From here we can easily write $v_+ = v(L)$ and $v_- = v(-L)$, giving
\begin{align} \label{eq const_full}
    v_{\pm} =& \pm \frac{\lambda}{2\eta}
    \Bigg[\left( \zeta -\frac{\gamma}{L} - k\frac{L-L^r}{L^r} + \frac{\lambda^2 L_c \zeta_i}{\lambda^2-L_c^2}\coth{(L/L_c)} - \frac{2\zeta\lambda^2}{4\lambda^2 - L_c^2}\left( 2 + \csch^2{(L/L_c)}\right) \right) \tanh{(L/\lambda)}
    \nonumber \\ & + \lambda L_c\left( \frac{2\zeta}{4\lambda^2-L_c^2}\coth{(L/L_c)}-\frac{\zeta_i L_c}{\lambda^2-L_c^2} \right) \Bigg], 
\end{align}
and $V = v_+ = -v_-$. The exact critical $L^*$ is such that \cref{eq const_full} = 0. In the relevant limit $L_c\ll L$ and $L_c\ll \lambda$, 
\begin{equation} \label{eq const_limitLc}
    v_{\pm} \approx \pm \frac{1}{2\eta}
    \left[ \lambda\tanh{(L/\lambda)}\left(L_c\zeta_i - \frac{\gamma}{L} - k\frac{L-L^r}{L^r} \right) + L_c\left( \frac{\zeta}{2} - \zeta_i L_c\right)\right], 
\end{equation}
and further, in the wet ($L \ll \lambda$) and dry ($L \gg \lambda$) cases,
\begin{align} \label{eq const_limits}
    v_{\pm}^{wet} &\approx \pm \frac{L_c}{2\eta}
    \left[ \zeta_i (L - L_c) + \frac{\zeta}{2}   \right]  \mp \frac{L}{2\eta}\left( \frac{\gamma}{L} + k\frac{L-L^r}{L^r} \right), \\ 
    v_{\pm}^{dry} &\approx \pm \frac{L_c}{2\eta}
    \left[ \zeta_i (\lambda - L_c) + \frac{\zeta}{2}   \right]  \mp \frac{\lambda}{2\eta}\left( \frac{\gamma}{L} + k\frac{L-L^r}{L^r} \right).
\end{align}
Setting $\gamma=0$ and $k=0$ and neglecting $L_c$ in front of $\lambda$ or $L$, we obtain \cref{eq const_wet} and \cref{eq const_dry} respectively.
\vfill

\section{Linear traction profile} \label{app grad_substrate}
For a linear traction profile $\zeta_i(x)$ (constant $\zeta_i'$), constant friction $\xi$ ($\xi'=0$), and same boundary conditions as in \cref{app const_substrate}, the solution to \cref{eq main} yields
\begin{align} \label{eq grad_full_profile}
    v(x) &= \frac{\lambda^2 L_c}{2\eta \sinh{(L/L_c)}} \Bigg[ \frac{\zeta}{4\lambda^2-L_c^2}\frac{\sinh{(2(x-X)/L_c)}}{\sinh{(L/L_c)}} - \frac{L_c \zeta_i(x)}{\lambda^2-L_c^2}\sinh{\Big(\frac{x-X}{L_c}\Big)}  +\frac{2\zeta_i'\lambda^2L_c^2}{(\lambda^2-L_c^2)^2}\cosh{\Big(\frac{x-X}{L_c}\Big)}\Bigg] \nonumber \\ & + C_1e^{x/\lambda}+C_2e^{-x/\lambda},   \text{  where  } \nonumber \\ 
    C_1 &=  \frac{-\lambda e^{-\frac{X}{\lambda}}}{4\eta \cosh{(L/\lambda)}}\Bigg[-\left( \zeta-\frac{\gamma}{L}-k\frac{L-L^r}{L^r}\right) + \lambda^2 L_c  \Bigg(
    \frac{2\zeta (1+\coth^2{(L/L_c)}) }{L_c(4\lambda^2-L_c^2)}\nonumber \\ & - \frac{\zeta_i(x)}{\lambda^2-L_c^2}\coth{(L/L_c)} 
    +  \frac{\zeta_i'\coth{(L/\lambda)}}{\lambda^2-L_c^2}\bigg( \frac{2\lambda^2L_c}{\lambda^2-L_c^2}-L_c-L\coth{(L/L_c)}\bigg) \Bigg) \Bigg], \nonumber \\ 
    C_2 &= \frac{\lambda e^{\frac{X}{\lambda}}}{4\eta \cosh{(L/\lambda)}} \Bigg[ -\left( \zeta-\frac{\gamma}{L}-k\frac{L-L^r}{L^r}\right) + \lambda^2 L_c  \Bigg(
    \frac{2\zeta (1+\coth^2{(L/L_c)}) }{L_c(4\lambda^2-L_c^2)} \nonumber \\ & - \frac{\zeta_i(x)}{\lambda^2-L_c^2}\coth{(L/L_c)} 
    +  \frac{\zeta_i'\coth{(L/\lambda)}}{\lambda^2-L_c^2}\bigg( \frac{2\lambda^2L_c}{\lambda^2-L_c^2}-L_c-L\coth{(L/L_c)}\bigg) \Bigg) \Bigg] \nonumber \\ &- \frac{\zeta_i'\lambda^3 L_c e^{\frac{X}{\lambda}}}{2\eta\sinh{(L/\lambda)}(\lambda^2-L_c^2)} \Bigg(\frac{2\lambda^2L_c}{\lambda^2-L_c^2} -L_c -L\coth{(L/L_c)}\Bigg) \Bigg]. 
\end{align}
The expressions for $v_+$ and $v_-$ are directly obtained by substituting $x_+ = X + L$ and $x_- = X - L$, giving 
\begin{align}
    v_\pm &= \pm \frac{\lambda^2 L_c}{2\eta} \Bigg[ \frac{2\zeta \coth{(L/L_c)}}{4\lambda^2-L_c^2} -  \frac{L_c \zeta_i^{\pm}}{\lambda^2-L_c^2} \pm \frac{2\zeta_i'\lambda^2L_c^2 \coth{(L/L_c)}}{(\lambda^2-L_c^2)^2}\Bigg] + C_1e^{\frac{X\pm L}{\lambda}}+C_2e^{-\frac{X\pm L}{\lambda}},  \label{eq grad_full_edges} \\
    U &= \frac{\zeta_i'}{2\eta}\frac{ L_c\lambda^2}{\lambda^2-L_c^2} \Big[  \lambda\coth{\left(\frac{L}{L_c}\right)} \left(\frac{2L_c^2 \lambda}{\lambda^2-L_c^2} + L\coth{\left(\frac{L}{\lambda}\right)}\right) - \frac{L_c \lambda (L_c^2 + \lambda^2 )}{\lambda^2-L_c^2}\coth{\left(\frac{L}{\lambda}\right)}  -L L_c\Big], \label{eq grad_full_v0}   \\
    V &= \frac{\lambda}{2\eta}
    \Bigg[\left( \zeta -\frac{\gamma}{L} - k\frac{L-L^r}{L^r} + \frac{\lambda^2 L_c \zeta_i(X)}{\lambda^2-L_c^2}\coth{(L/L_c)} - \frac{2\zeta\lambda^2}{4\lambda^2 - L_c^2}\left( 2 + \csch^2{(L/L_c)}\right) \right) \tanh{(L/\lambda)}
    \nonumber \\ & + \lambda L_c\left( \frac{2\zeta}{4\lambda^2-L_c^2}\coth{(L/L_c)}-\frac{\zeta_i(X) L_c}{\lambda^2-L_c^2} \right) \Bigg], \label{eq grad_full_vs}  
\end{align}  
which is equal to \cref{eq const_full} from the uniform case, with $\zeta_i = \zeta_i(X)$. Importantly, $U$ does not depend on the traction offset $\zeta_i^0=\zeta_i(X),\zeta,\gamma$ or $k$. In the relevant limit $L_c\ll \lambda$ and $L_c\ll L$,
\begin{equation}  \label{eq grad_limitLc}
    v_\pm \approx \pm \frac{L_c}{2\eta} \Bigg[ \frac{\zeta}{2} - L_c \zeta_i^{\pm} \pm 2\zeta_i' L_c^2 \Bigg] \mp \frac{\lambda}{2\eta} \Bigg[\Big(\frac{\gamma}{L} + k\frac{L-L^r}{L^r} - L_c\zeta_i(X) \Big)\tanh{\left(\frac{L}{\lambda}\right)}  \mp L_c L\zeta_i'\coth{\left(\frac{L}{\lambda}\right)}\Bigg]
\end{equation}

In the wet ($L \ll \lambda$) and dry ($L \gg \lambda$) cases,
\begin{align} \label{eq grad_limits}
    v_\pm^{wet} &\approx \pm \frac{L_c}{2\eta}\Bigg[ L\zeta_i^{\pm} + \frac{\zeta}{2} \pm \zeta_i'\left(\lambda^2 - \frac{2}{3}L^2\right) \Bigg]   \mp  \frac{L}{2\eta} \Big(\frac{\gamma}{L} + k\frac{L-L^r}{L^r} \Big)  \\ 
    v_\pm^{dry} &\approx \pm \frac{L_c}{2\eta}\Bigg[ \lambda \zeta_i^{\pm} + \frac{\zeta}{2} \pm 2\zeta_i'L_c^2 \Bigg] \mp \frac{\lambda}{2\eta} \Big(\frac{\gamma}{L} + k\frac{L-L^r}{L^r} \Big). 
\end{align}
Setting $\gamma=0$ and $k=0$ we obtain \crefrange{eq grad_dry}{eq grad_wetv0}.
\vfill

\newpage
\section{Edge velocities} \label{app v0vs_edges} 
Here we present some examples of plots including both spreading and center-of-mass velocity, on the one hand, and edge velocities on the other hand, for better illustration and understanding of the phenomenology.  

\begin{figure}[htb!]
\centering
\begin{minipage}{.45\textwidth}
  \centering
  \includegraphics[width=0.87\linewidth, trim={0cm 1cm 0 0},clip]{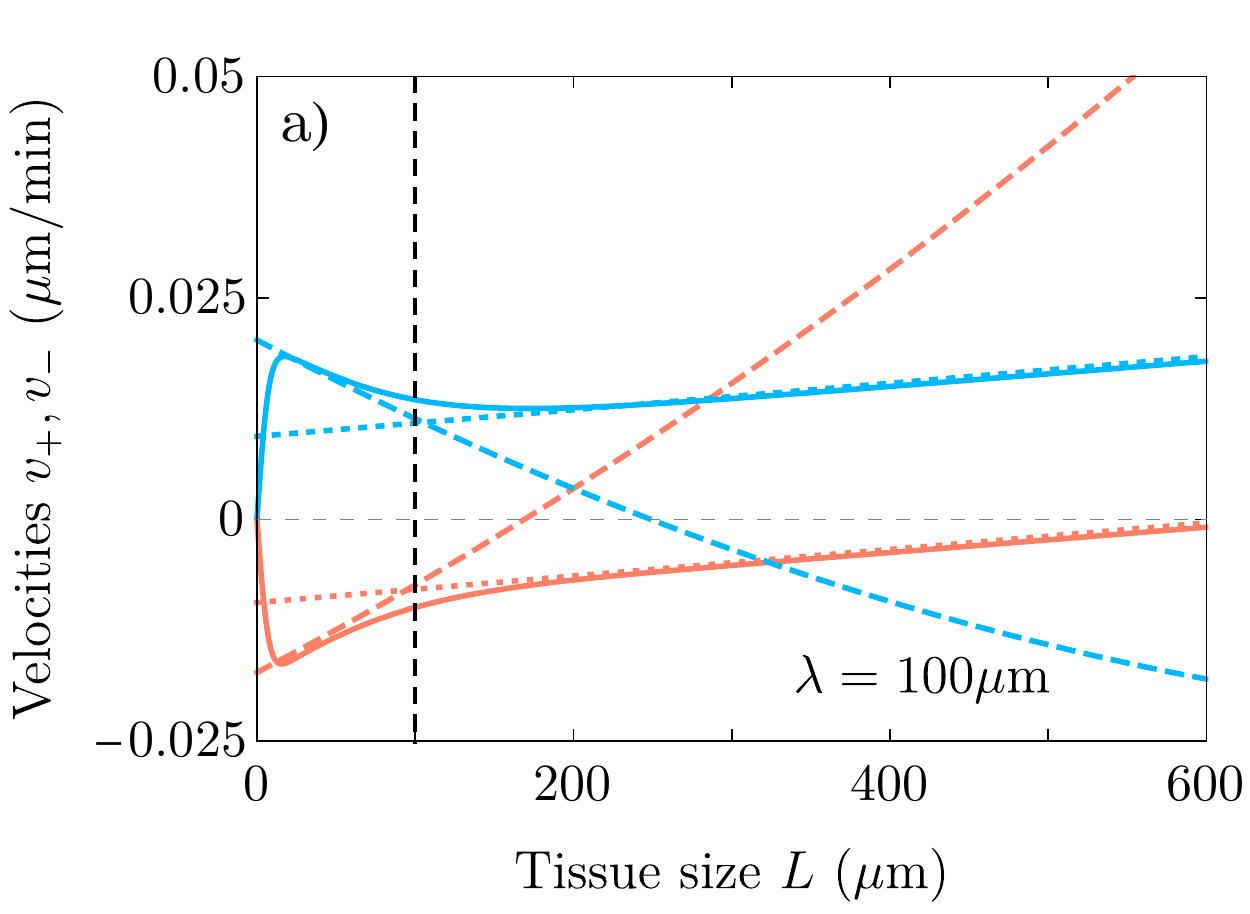} \vfill
  \includegraphics[width=0.87\linewidth, trim={0cm 1cm 0 0},clip]{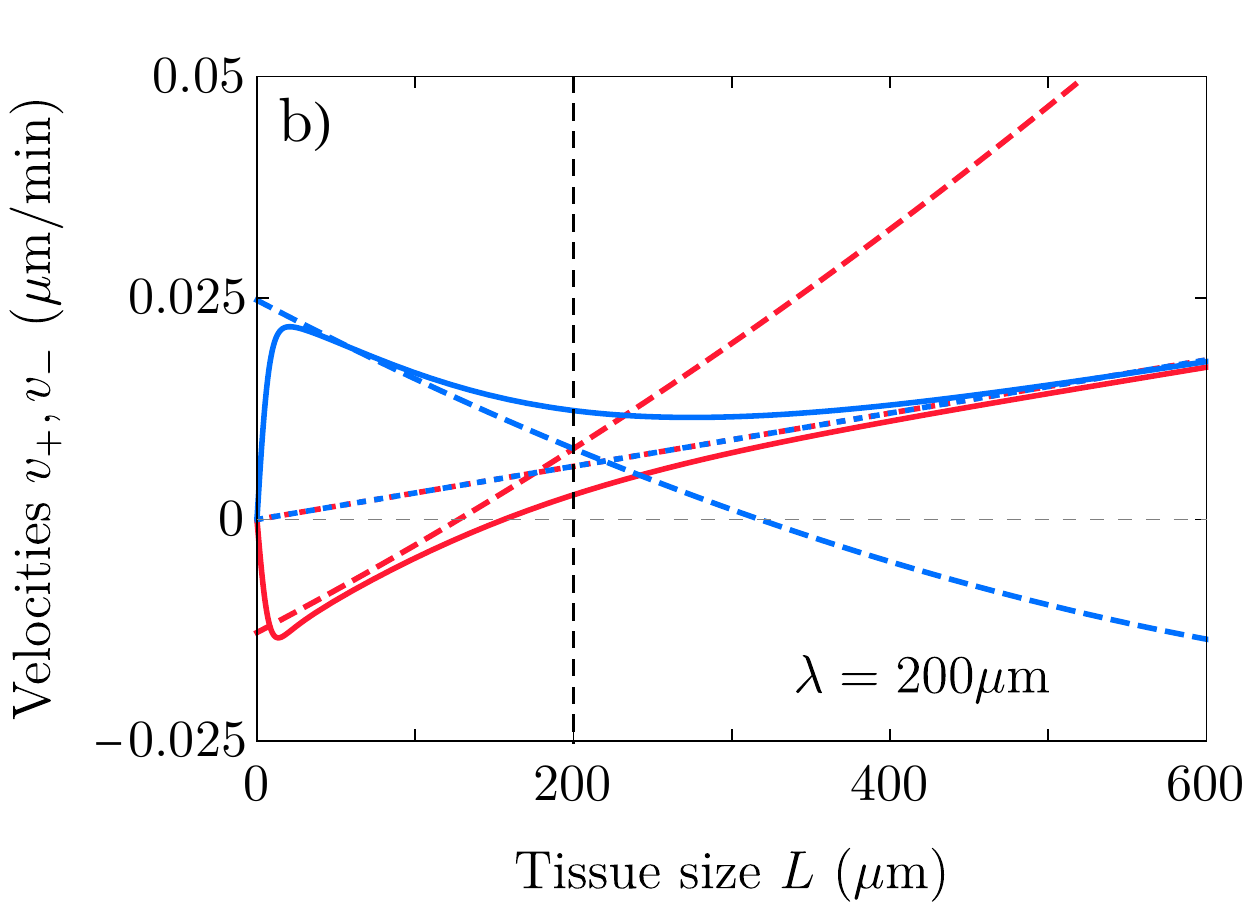} \vfill
  \includegraphics[width=0.87\linewidth]{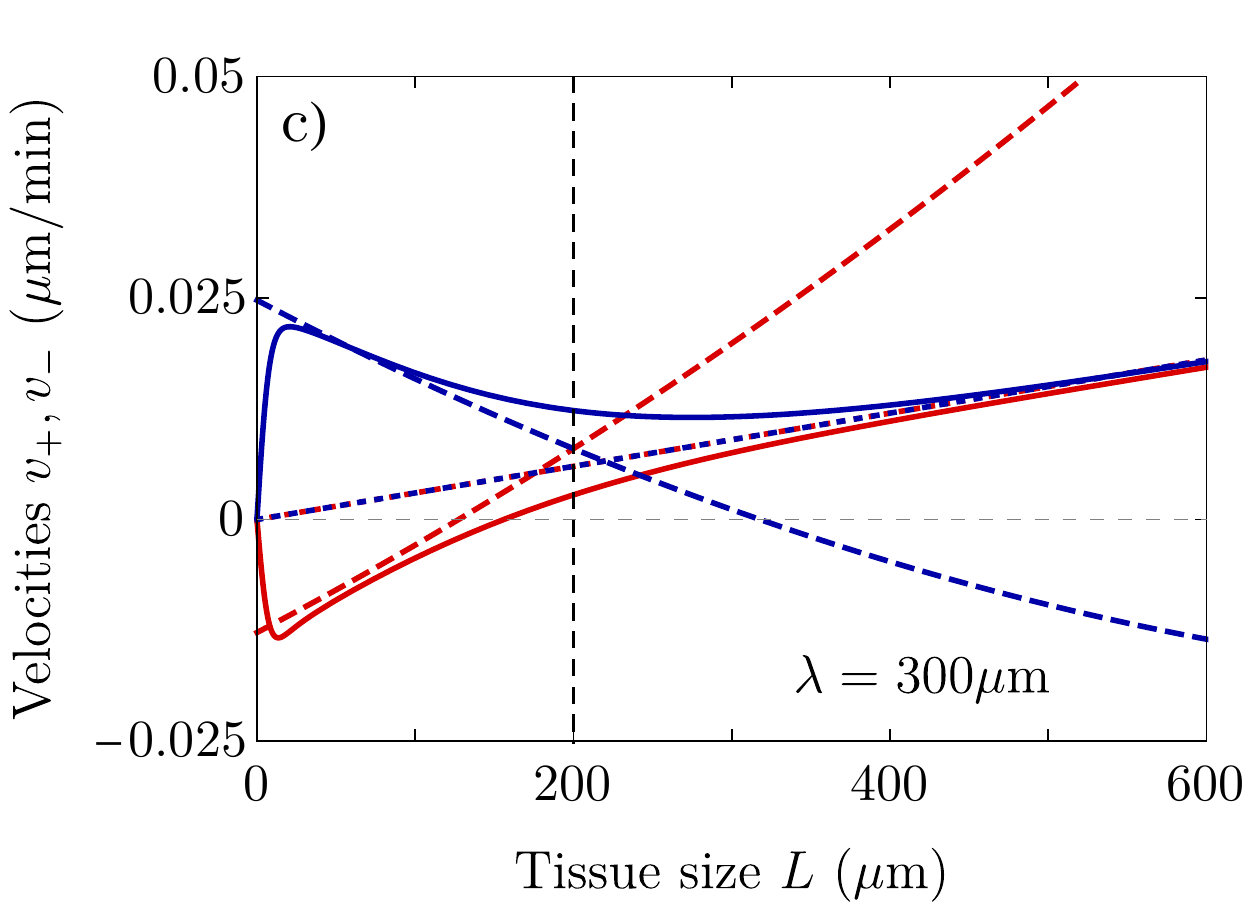}
  \captionof{figure}{Edge velocities $v_-$ (blue lines) and $v_+$ (red lines) in their full expressions (continuous), dry (dotted) and wet (dashed) limits, for three different values of $\lambda$ (vertical dashed lines) and a constant $L_p=200$ $\mu$m. Equivalent to the  examples giving $V$ and $U$ in \cref{fig v0vs_lambdacases}.}
  \label{fig v0vs_lambdacases_edges}
\end{minipage} \hspace{1cm} 
\begin{minipage}{.45\textwidth}
  \centering
  \includegraphics[width=0.85\linewidth, trim={0cm 1cm 0 0},clip]{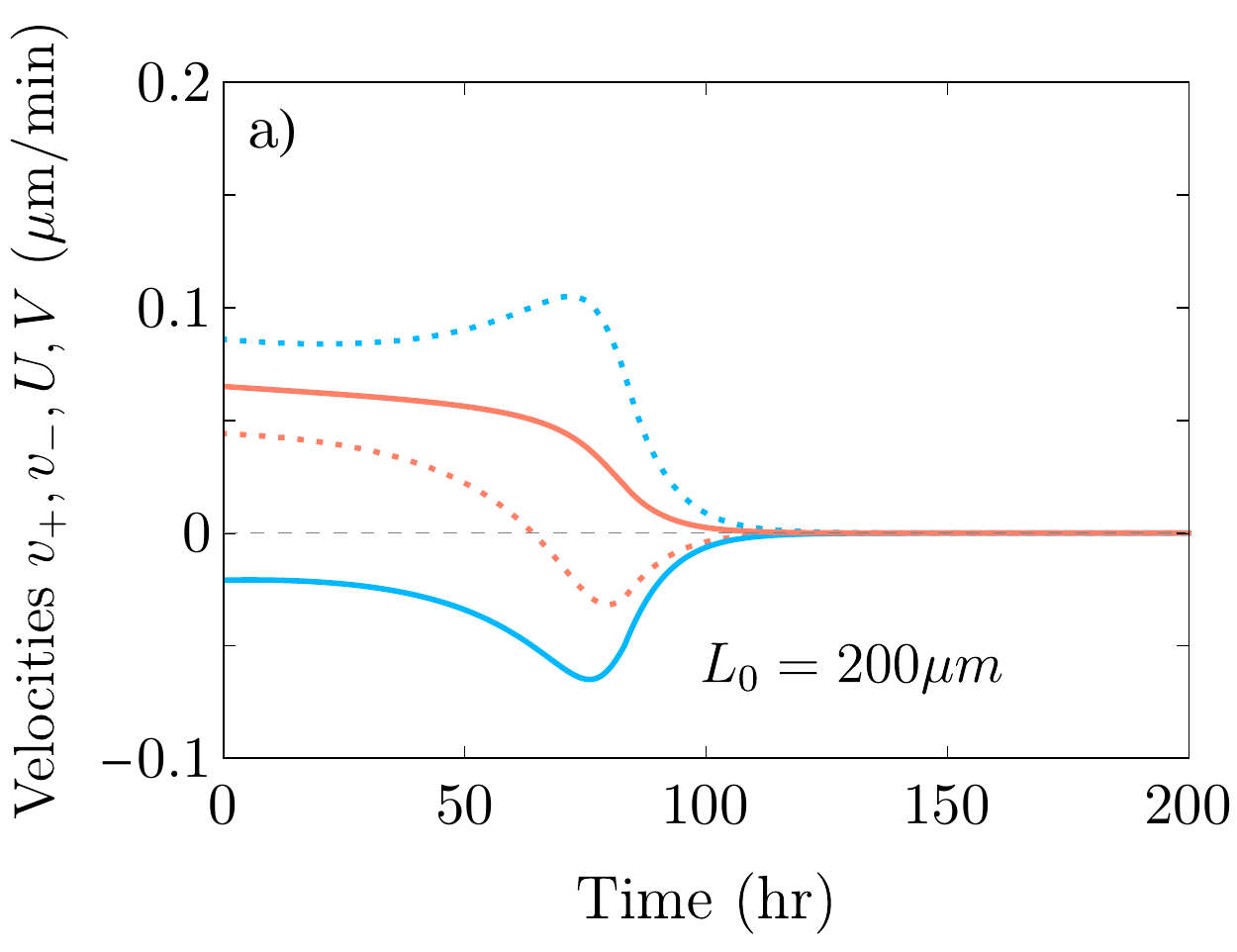} \vfill
  \includegraphics[width=0.85\linewidth, trim={0cm 1cm 0 0},clip]{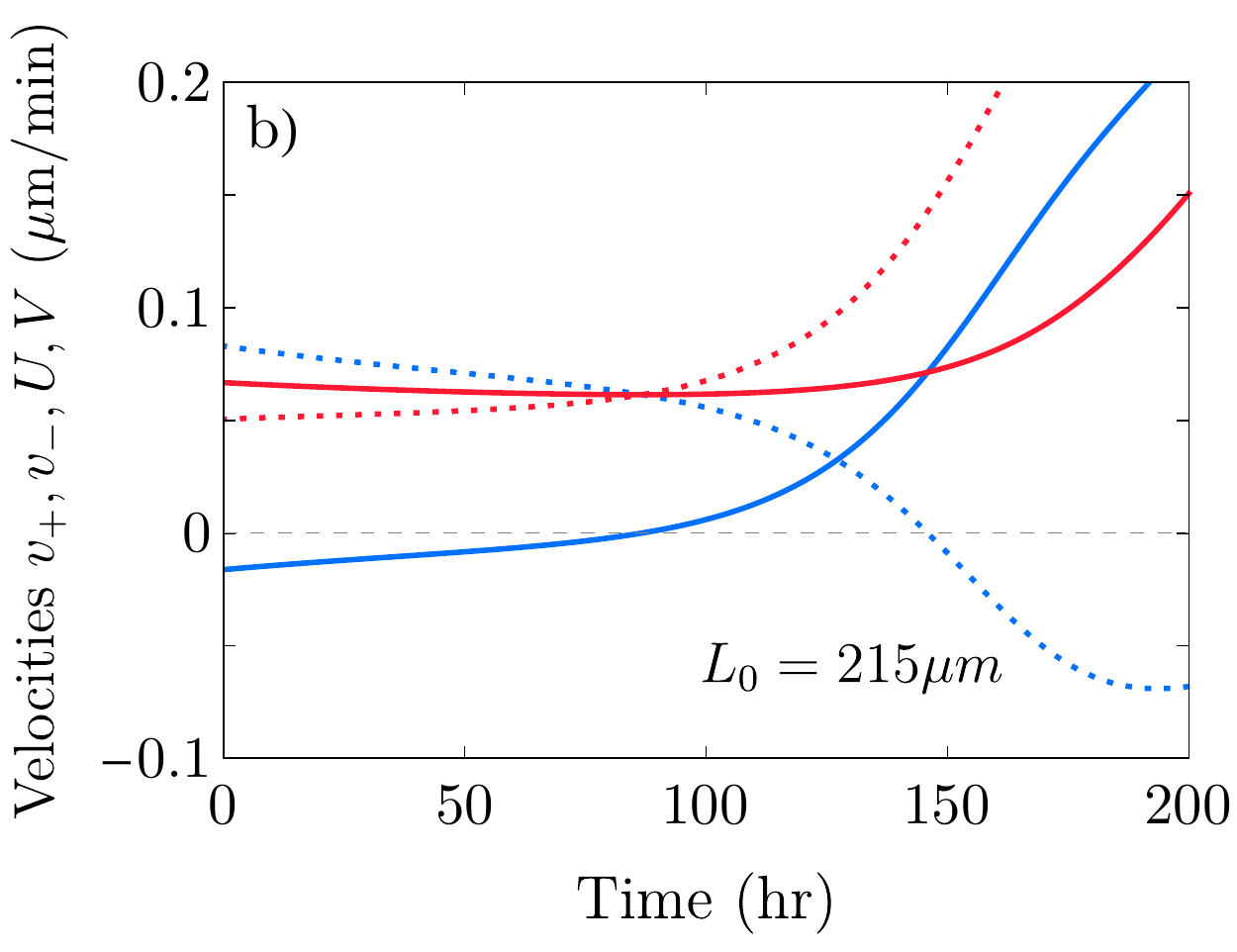} \vfill
  \includegraphics[width=0.85\linewidth]{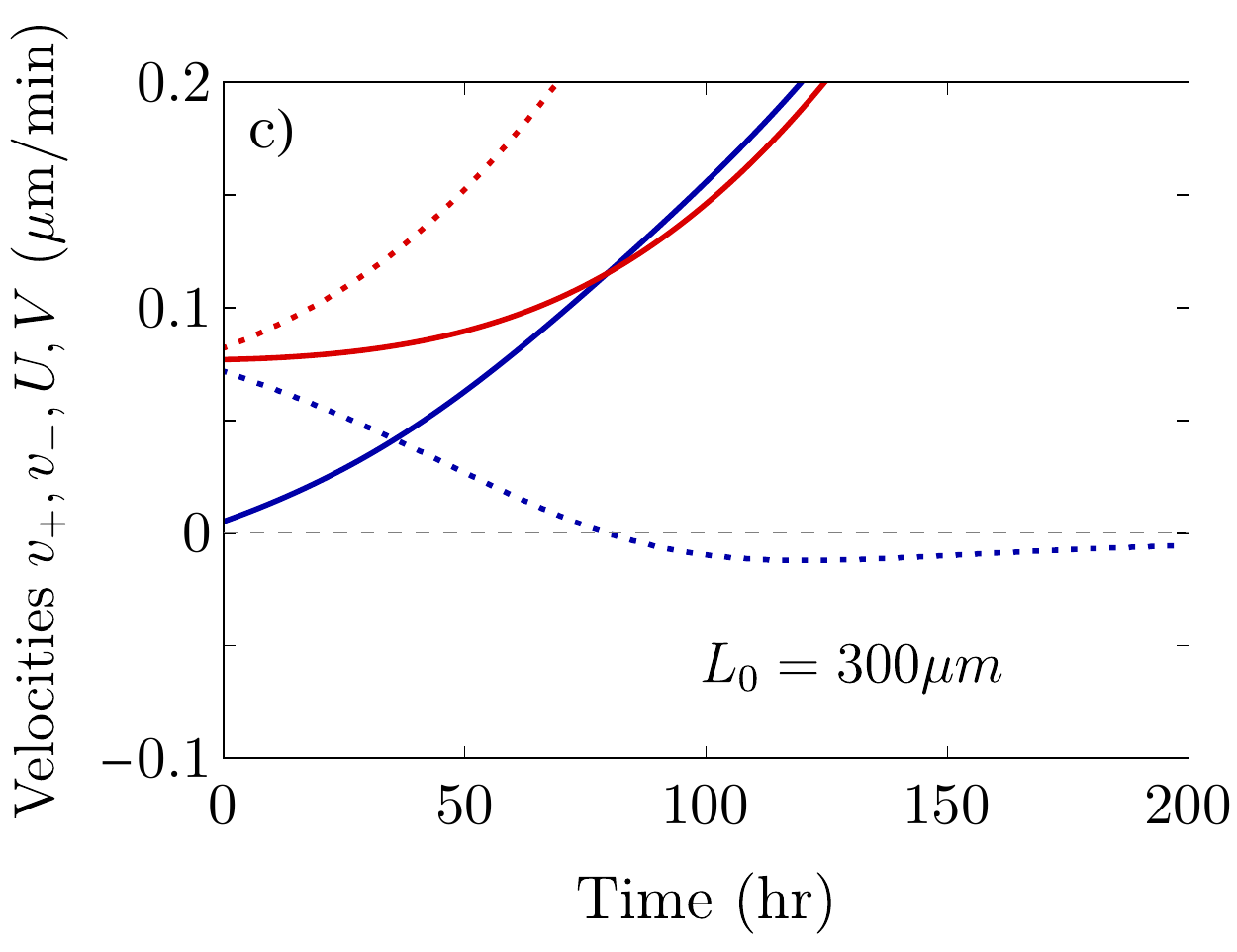}
  \captionof{figure}{Time evolution of the $v_-$ (blue dotted), $v_+$ (red dotted), $V$ (blue) and $U$ (red) velocities, for clusters starting in three different values of $L_0$, characteristic of the three regimes. They correspond to the same examples from \cref{fig evolution}.}
  \label{fig  evolution_edges}
\end{minipage}
\end{figure}

\end{document}